\documentclass[letterpaper,journal]{IEEEtran}
\usepackage{amsmath,amsfonts}
\usepackage{algorithmic}
\usepackage{algorithm}
\usepackage{amssymb}
\usepackage{array}
\usepackage[caption=false,font=normalsize,labelfont=sf,textfont=sf]{subfig}
\usepackage{textcomp}
\usepackage{stfloats}
\usepackage{url}
\usepackage{verbatim}
\usepackage{graphicx}
\usepackage{svg}
\usepackage{booktabs}
\usepackage{cite}
\usepackage[colorlinks=true, linkcolor=black, citecolor=black, urlcolor=black]{hyperref}

\makeatletter
\long\def\@makecaption#1#2{%
  \vskip\abovecaptionskip
  \footnotesize                 
  \sbox\@tempboxa{#1. #2}
  \ifdim \wd\@tempboxa >\hsize
    #1. #2\par                  
  \else
    \global\@minipagefalse
    \hb@xt@\hsize{\hfil\box\@tempboxa\hfil}
  \fi
  \vskip\belowcaptionskip}
\makeatother

\begin{document}
\title{Learned Digital Codes for Over-the-Air Computation in Federated Edge Learning}

\author{Antonio Tarizzo, Mohammad Kazemi, Deniz G\"und\"uz
\thanks{M. Kazemi’s work was funded by UK Research and Innovation (UKRI) under the UK government’s Horizon Europe funding guarantee [grant number 101103430].}
\thanks{The authors are with the Department of Electrical and Electronic
Engineering, Imperial College London, SW7 2BT, London, U.K. (emails:
{antonio.tarizzo24,mohammad.kazemi,d.gunduz}@imperial.ac.uk).}}

\maketitle

\begin{abstract}
Federated edge learning (FEEL) enables wireless devices to collaboratively train a centralised model without sharing raw data, but repeated uplink transmission of model updates makes communication the dominant bottleneck. Over-the-air (OTA) aggregation alleviates this by exploiting the superposition property of the wireless channel, enabling simultaneous transmission and merging communication with computation. Digital OTA schemes extend this principle by incorporating the robustness of conventional digital communication, but current designs remain limited in low signal-to-noise ratio (SNR) regimes. This work proposes a learned digital OTA framework that improves recovery accuracy, convergence behaviour, and robustness to challenging SNR conditions while maintaining the same uplink overhead as state-of-the-art methods. The design integrates an unsourced random access (URA) codebook with vector quantisation and AMP-DA-Net, an unrolled approximate message passing (AMP)-style decoder trained end-to-end with the digital codebook and parameter server local training statistics. The proposed design extends OTA aggregation beyond averaging to a broad class of symmetric functions, including trimmed means and majority-based rules. Experiments on highly heterogeneous device datasets and varying numbers of active devices show that the proposed design extends reliable digital OTA operation by more than 10 dB into low SNR regimes while matching or improving performance across the full SNR range. The learned decoder remains effective under message corruption and nonlinear aggregation, highlighting the broader potential of end-to-end learned design for digital OTA communication in FEEL.
\end{abstract} 

\begin{IEEEkeywords}
Federated edge learning, digital over-the-air computation, unsourced random access, distributed optimisation, compressed sensing.
\end{IEEEkeywords} 

\section{Introduction}
\IEEEPARstart{F}{ederated} learning (FL) has become a well-established approach in machine learning (ML), enabling multiple clients to collaboratively train a shared model while keeping data local, addressing privacy concerns, and reducing reliance on centralised storage\cite{mcmahan2017communication, konecny2016federated_strategies}. This interest naturally extends to wireless edge devices such as smartphones, Internet of things (IoT) sensors, and autonomous systems, which generate large volumes of data that are highly valuable for training ML models\cite{jia2025comprehensive_survey}. In the context of edge ML, FL also reduces bandwidth usage by avoiding the transmission of raw data over constrained wireless links. However, scaling FL to the wireless edge remains challenging as model dimensions grow \cite{Chen:JSAC:21}. Each device must repeatedly upload model updates to the server, creating a significant bottleneck on the wireless uplink. While compression techniques such as sparsification\cite{aji2017sparse, Ozfatura:ISIT:21} and quantisation\cite{alistarh2017qsgd, bernstein2018signSGD, seide2014_onebit_sgd, isik2023sparserandomnetworks} help reduce payload size, there is also an opportunity to optimise how communication itself is performed, enabling the aggregation operation to be partially or fully merged with the wireless transmission process.

Over-the-air (OTA) aggregation addresses the wireless uplink bottleneck by allowing devices to transmit simultaneously, exploiting the linear superposition property of the wireless channel to combine updates directly \cite{Amiri:TSP:20, Amiri:TWC:20}. This removes scheduling overhead and significantly reduces uplink latency. One line of OTA research operates in the analog domain, where devices transmit uncoded updates that are aggregated over the air. Although these methods can be highly efficient, they are sensitive to noise, fading, and transmit power misalignment, which makes robust deployment in practical wireless environments challenging \cite{jia2025comprehensive_survey, sun2022_dynamic_scheduling_ota_fe_edge_learning}. Digital OTA provides an alternative by encoding model updates into discrete sequences prior to transmission. This retains the ability to aggregate updates over the air while benefiting from the robustness, modularity, and compatibility of conventional digital communication systems \cite{zhu2021onebit_over_the_air_aggregation}. However, existing digital OTA designs, such as massive digital over-the-air computation (MD-AirComp), either show poor convergence or degrade sharply in the low signal-to-noise ratio (SNR) regime typical of IoT devices and are largely limited to simple averaging objectives \cite{qiao2024_massive_digital_ota_computation}. These limitations motivate the development of more robust digital OTA frameworks in which learning-based decoding and codebook design enable improved recovery and convergence without increasing uplink overhead.

\textbf{Related Works:}
Early implementations of federated edge learning (FEEL) relied on orthogonal multiple access schemes such as time division multiple access (TDMA) and orthogonal frequency division multiple access (OFDMA), where devices are allocated distinct channel resources. While reliable and straightforward, these approaches incur high latency since each update must be decoded individually before aggregation. This issue is addressed in \cite{zhu2019_broadband_analog_aggregation} by proposing broadband analog aggregation, where devices scale their model updates to partially invert the channel and transmit them simultaneously so that the server directly observes a noisy sum on each subcarrier. In \cite{yang2020_fed_ota}, OTA computation is subsequently applied to FEEL with a multi-antenna server, jointly optimising device selection and receive beamforming to satisfy a mean-squared error (MSE) constraint while maximising the number of participating devices. These analog OTA schemes demonstrate substantial uplink savings, but their performance is limited in practice by sensitivity to noise and fading in the low SNR regime.

One-bit digital OTA aggregation (OBDA) was an early step towards digital designs, replacing full-precision gradients with their signs and mapping each sign to a one-bit symbol, which all devices transmitted simultaneously~\cite{zhu2021onebit_over_the_air_aggregation}. The receiver effectively performs a noisy majority vote over these signs and applies the approach in \cite{bernstein2018signSGD} for optimisation, which has been shown to converge, but with a noticeable loss in final accuracy due to the extreme quantisation. The frequency-shift keying with majority vote (FSK-MV) scheme in \cite{sahin2021_distributed_learning_fsk_majority_vote} improves robustness to hardware imperfections and non-coherent reception by encoding gradient signs onto orthogonal tones and deciding the global sign via an energy-based majority vote. However, it still conveys essentially one bit per update dimension and therefore faces a similar accuracy ceiling.

Massive digital over-the-air computation (MD-AirComp) extends digital OTA to the FEEL setting with large numbers of intermittently active devices by combining ideas from compressed sensing, vector quantisation~\cite{qiao2024_massive_digital_ota_computation}, and unsourced random access (URA)~\cite{polyanskiy2017_perspective_massive_random_access,URA_Survey}, where a common codebook is used by all devices since the server is only interested in a function of set of messages transmitted by devices (e.g., their average), not the individual messages. Note that incorporating URA makes the solution scalable with the number of devices \cite{URA_Survey}.
Next, each device quantises its local model update using a shared codebook, then maps the resulting indices to a sequence of URA codewords. All devices transmit these codewords simultaneously, and the server applies an approximate message passing (AMP)-based decoder to estimate the frequency with which each codeword was selected, which is then mapped to an approximate arithmetic mean of the model updates. MD-AirComp achieves higher accuracy than analog OTA and greater efficiency than orthogonal access. Still, in the low-SNR regime, the sparse recovery and activity estimation become unreliable, and global training fails to converge. This gap motivates the development of learned digital OTA schemes that retain MD-AirComp’s scalability and compatibility with URA while improving robustness, convergence, and flexibility across different aggregation functions. 

Beyond FEEL-specific schemes, digital OTA has also been studied for more general function computations over a wireless multiple-access channel. ChannelComp \cite{razavikia2024_channelcomp} formulates digital OTA as the problem of choosing transmit constellations so that, after superposition, the received symbol reveals the value of a given finite-valued function, which can then be read out via a simple lookup at the receiver. SumComp \cite{razavikia2025_sumcomp} specialises this idea to sums and means by using ring-of-integers mappings on quadrature amplitude modulation (QAM) and pulse amplitude modulation (PAM) constellations, leading to low-complexity encoders and improved MSE for arithmetic aggregation. The bit-slicing approach in \cite{liu2025_bit_slicing} improves reliability by spreading the bits of each quantised value across multiple channel uses and protecting high-significance bits more strongly at detection, while \cite{xie2023_joint_coding_modulation} proposes a joint channel-coding and modulation design based on non-binary low-density parity-check (LDPC) codes tailored to digital OTA operation. These schemes broaden the class of functions that can be computed and strengthen physical-layer robustness, but typically assume per-device constellations and fixed user sets, and do not explicitly account for federated training dynamics or massive, dynamic device activities.

A complementary line of work studies function computation over wireless channels through nomographic representations, in which a multivariate function can be written as
\begin{equation}
        f(x_1,\ldots,x_d)
    \;=\;
    \psi \left( \sum_{i=1}^d \phi_i(x_i) \right),
    \label{nomo}
\end{equation}
with device-side pre-processing functions $\phi_i(\cdot)$ and a server-side post-processing function $\psi(\cdot)$. This structure naturally aligns with the additive multiple-access channel (MAC), as the channel computes the sum over the air. Building on this idea, the authors in \cite{bouzinis2024_universal_ota_function_approximation} employ deep neural networks to learn $\phi_i$ and $\psi$ in a centralised manner and then split the model into device-side and server-side subnetworks, enabling universal function approximation over MAC. While the focus in~\cite{bouzinis2024_universal_ota_function_approximation} is on learning general pre- and post-processing mappings for analog function computation, the proposed digital OTA framework applies learning directly to the digital communication layer. It is important to note that our proposed method remains compatible with nomographic computation by allowing device-side pre-processing before transmission and server-side post-processing after digital decoding

Classical AMP-based digital OTA aggregation schemes rely on analytically derived priors and asymptotic assumptions that are systematically violated in practical FEEL settings due to finite codebooks, structured URA designs, non-uniform codeword popularity, quantisation mismatch, and strong SNR variability. While existing learning-based FEEL designs have applied neural networks to higher-layer tasks such as power control and client scheduling in digital OTA FL, or to device-side pre-processing and server-side post-processing in analog OTA computation, learning has not previously been applied directly to the digital OTA aggregation pipeline. Rather than proposing a new message-passing algorithm, this work integrates learning into the digital OTA pipeline by jointly training the decoder and URA codebook. This data-driven design enables adaptive residual scaling, damping, and prior modelling from observed training dynamics, resulting in reliable sparse recovery and stable global convergence in communication regimes where analytically designed AMP-based methods fail.

Our main contributions in this paper are:

\noindent \textit{$\bullet$ Improved performance and generalisation:} The proposed framework extends reliable digital OTA FEEL operation by more than $10$ dB in lower SNRs compared to the state-of-the-art, MD-AirComp, while maintaining or improving performance across the full SNR range at the same uplink overhead. Moreover, the learned decoder generalises effectively across different global model architectures, varying numbers of active devices, and when devices' data highly deviates from independent and identically distributed (IID) distributions.

\noindent \textit{$\bullet$ Generalised digital OTA aggregation:} The proposed framework extends beyond averaging-only aggregation to a general class of symmetric functions, including majority voting, trimmed means, and other robust statistics. This is enabled by a separable digital OTA design in which individual device messages can be reconstructed at the server and then passed through an arbitrary symmetric rule. We evaluate this flexibility in scenarios with device message corruption and systematic bias, illustrating robustness beyond simple mean aggregation. 

\noindent \textit{$\bullet$ End-to-end learned decoder and quantisation:} We introduce AMP-DA-Net, an unrolled AMP-style decoder that is trained end-to-end together with the URA-based digital encoder and vector quantiser. The design incorporates codeword popularity priors and a \textit{simple VQ} (SimVQ)-inspired codebook parameterisation. Moreover, we further study curvature-aware vector quantisation and a quantisation-aware loss that encourages low distortion between ideal and quantised updates. This learned communication layer consistently outperforms fixed AMP and hand-crafted codebooks under the same uplink budget.

\noindent \textit{$\bullet$ Broader implications for learned digital OTA:}
Our results show that jointly learning the URA and quantisation codebooks, along with the decoder, can substantially improve robustness and accuracy relative to fixed constructions. This suggests a broader role for end-to-end learned designs in URA-based access schemes and digital OTA systems beyond the specific FEEL setting considered here, and provides a flexible framework for future exploration of more general function learning and richer system constraints.

\textit{Relation to prior conference version:} A preliminary version of this work, focusing on fixed aggregation and limited system configurations, was submitted to ICASSP 2026 \cite{tarizzo2025_learnt_digital_codes}. The present journal version substantially extends that work by including quantisation into the end-to-end learned pipeline, enabling generalised aggregation functions, testing robustness to corrupted devices, and a comprehensive system-level analysis across diverse communication and learning regimes.

\textbf{Paper Structure:} The remainder of this paper is organised as follows. Section II describes the system model and formulates the symmetric aggregation problem. Section III presents the proposed generalised digital OTA framework, including the construction of the quantisation codebook, the learned URA codebook, and the decoder design. Section IV reports the main results and comparisons with baselines and the state-of-the-art scheme. Section V provides an extensive evaluation that covers variations of the proposed method, a generalisation study, corruption scenarios that demonstrate alternative aggregation methods, and codebook construction. Section VI concludes the paper and highlights directions for future work.

\textbf{Notation:} lower-case, bold lower-case, and bold upper-case denote scalars, vectors, and matrices, respectively. $\mathbf{I}_l$ denotes the $l\times l$ identity matrix. $\mathbf{0}$ and $\mathbf{1}$ denote all-zero and all-one vectors of appropriate dimension. For a vector $\mathbf{x}$, $x_i$ denotes its $i$-th entry. For a matrix $\mathbf{X}$, $X_{ij}$ denotes its $(i,j)$-th entry and $\mathbf{X}_j$ its $j$-th column. The $p$-norm is denoted by $\|\cdot\|_p$. The operators $\odot$ and $\oslash$ denote element-wise multiplication and division. The operator $\circ$ denotes element-wise exponentiation, e.g., $\mathbf{C}^{\odot 2}$ is the element-wise square of $\mathbf{C}$. Superscripts in parentheses, e.g.\ $(\ell)$, denote the network layer or iteration index of an algorithm. Expectation, variance, and covariance are denoted by $\mathbb{E}[\cdot]$, $\mathrm{Var}(\cdot)$, and $\mathrm{Cov}(\cdot,\cdot)$, respectively. The Dirac measure at zero is denoted by $\delta_0$. Set-builder notation $\{x_k : k \in \mathcal{S}\}$ denotes the unordered set of elements $x_k$ for indices $k$ in set $\mathcal{S}$. Moreover, $(\cdot)^\top$ and $|\cdot|$ denote transpose and cardinality operators, respectively. The sets of real numbers and non-negative integers are denoted by $\mathbb{R}$ and $\mathbb{N}_0$, respectively.

\section{System Model} \label{sec:system}
\begin{figure*}[t]
    \centering
    \includegraphics[width=0.84\textwidth]{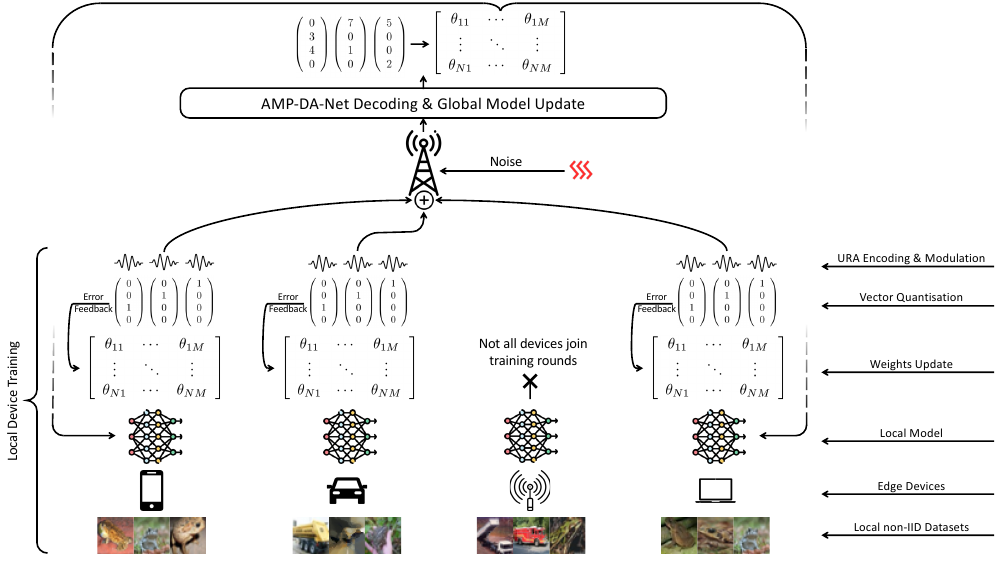}
    \caption{Illustrative digital OTA pipeline for FEEL.}
    \vspace{-10pt}
    \label{fig:FEEL_art}
\end{figure*}

We consider a FEEL system with $K_t$ edge devices, of which an unknown random subset $\mathcal{S}_a$ with $|\mathcal{S}_a| = K_a$ is active in each round. This models how only a subset of devices participate in any given round due to heterogeneous computation speeds, energy and connectivity constraints, and varying data availability. With a fixed aggregation deadline as in this scheme, only devices that finish local training in time and opt to transmit become active, leading to $K_a \ll K_t$ in most practical scenarios~\cite{qiao2024_massive_digital_ota_computation,li2020_fedprox,khajehali2023_iot_fl_survey}. Furthermore, each device has a single antenna and maintains a local dataset $\mathcal{D}_k$. For convenience, devices are assumed to have equal-sized datasets, $|\mathcal{D}_k| = D$, and equal computational capacity. Although heterogeneity affects participation dynamics in practice, it does not alter the underlying communication model and can therefore be abstracted away~\cite{li2020_fedprox}. A single-antenna base station (BS) serves as both receiver and parameter server, handling uplink decoding, aggregation, and global model updates. The BS also maintains a local dataset, $\mathcal{D}_0$, which it uses for local training to construct the quantisation codebook in each round (see Section~\ref{quant_codebook}), and may optionally be included in the global aggregation depending on the quality and relevance of $\mathcal{D}_0$. Moreover, symbol-level synchronisation is assumed so that transmissions from devices arrive at the BS almost simultaneously, i.e., during a single symbol duration, as in most OTA-FL models~\cite{qiao2024_massive_digital_ota_computation,Sahin:PIMRC:25}. An overview of the considered system model is depicted in Fig. \ref{fig:FEEL_art}.

\subsection{Learning Objective}
The standard federated learning objective is to minimise the aggregate loss function $f(\mathbf{w})$, where $\mathbf{w}\in\mathbb{R}^W$ denotes the vector of global model parameters, with $W$ being the number of parameters. 
The main objective can be written as
\begin{equation}
    \min_{\mathbf{w}\in\mathbb{R}^W} 
    f(\mathbf{w})
    ~=~
    \frac{1}{K_t}\sum_{k=1}^{K_t} F_k(\mathbf{w}),
    \label{eq:erm_standard}
\end{equation}
where $F_k(\mathbf{w})$ is the empirical loss of this model on the local dataset $\mathcal{D}_k$. At the $t$-th global iteration, each device starts from the current global model $\mathbf{w}^t$, performs $E$ steps of local stochastic gradient descent (SGD) on $F_k$, obtain the updated local parameters, $\mathbf{w}_k^t$, and sends the difference in model parameters, $\mathbf{w}_k^t-\mathbf{w}^t$, to the BS. Under the objective in~\eqref{eq:erm_standard}, the gradient of $f(\mathbf{w})$ is proportional to the arithmetic mean of the local gradients, so averaging these updates is a natural aggregation rule. 

Note that in some practical scenarios, imperfections at the server or devices, e.g., due to hardware impairments~\cite{HWI,URA_HWI}, can introduce systematic biases that the arithmetic mean cannot remove when used as the aggregation rule. In such settings, alternative aggregation methods such as majority voting, trimmed means, or other robust statistics may be preferred. This motivates a more general formulation, similar to the nomographic formulation in~\eqref{nomo}, in which each active device transmits a message $\mathbf{m}_k^t$, derived from its local update after $E$ SGD steps, via a pre-processing operation. The BS then recovers estimates of these transmitted messages and applies a \emph{symmetric aggregation function} (post-processing function), $g(\cdot)$, to obtain an aggregate update vector $\mathbf{g}^t$. That is,
\begin{equation}
    \mathbf{g}^t
    ~=~
    g\big(\{\hat{\mathbf{m}}_k^t : k\in\mathcal{S}_a\}\big),
    \label{eq:sym_agg}
\end{equation}
where $\hat{\mathbf{m}}_k^t$ denotes the recovered estimate of the transmitted message $\mathbf{m}_k^t$ obtained after uplink decoding. The function $g(\cdot)$ is permutation-invariant and depends only on the multiset of recovered messages, not on their ordering. The arithmetic mean is recovered as a special case, but~\eqref{eq:sym_agg} allows more general nonlinear, discrete-valued, or robust aggregation rules such as majority voting, trimmed mean, or learned symmetric mappings tailored to the communication setting.

\subsection{Communication Model}
At the start of round $t$, the BS broadcasts the updated global model parameters $\mathbf{w}^{t-1}$ to all devices. We follow the common assumption in FEEL and OTA-FL that the downlink is sufficiently reliable and well-provisioned (e.g., via broadcast or coded multicast~\cite{bonawitz2019_towards_fl_scale}), so its cost is not modelled explicitly, and the dominant communication bottleneck lies on the uplink~\cite{Chen:JSAC:21}. The $k$-th active device then performs $E$ local SGD steps to produce $\mathbf{w}_k^t$, generating its local update
\begin{equation}
    \Delta \mathbf{w}_k^t = \mathbf{w}_k^t - \mathbf{w}^{t-1}.
\end{equation}
Devices can also employ error-feedback quantisation to account for quantisation errors over time. That is, the $k$-th active device maintains an error accumulator, $\mathbf{e}_k^t$, updated as
\begin{align}
    \mathbf{s}_k^t &= \Delta \mathbf{w}_k^t + \mathbf{e}_k^{t-1}, \label{s_update}\\
    \mathbf{e}_k^t &= \mathbf{s}_k^t - \mathcal{Q}(\mathbf{s}_k^t),
\end{align}
where $\mathcal{Q}(\cdot)$ denotes the quantisation operator, $\mathcal{Q}(\mathbf{s}_k^t)$ is the quantised message transmitted over the uplink and $\widehat{\mathcal{Q}}(\mathbf{s}_k^t)$ is the BS estimate after uplink decoding. The BS receives these compressed messages and applies the symmetric aggregation rule in \eqref{eq:sym_agg} to obtain
\begin{equation}
    \mathbf{g}^t
    ~=~ g\big(\{\widehat{\mathcal{Q}}(\mathbf{s}_k^t) : k\in\mathcal{S}_a\}\big).
    \label{g}
\end{equation}
Finally, the global model is updated with a global learning rate $\eta$, as 
\begin{equation}
    \mathbf{w}^{t}
    ~=~ \mathbf{w}^{t-1} + \eta\,\mathbf{g}^t.
    \label{update}
\end{equation}

\section{Proposed Solution}
\label{sec:proposed}
The proposed framework integrates a jointly learned encoder-decoder design into the FEEL uplink. To train the communication layer, we first generate an offline training dataset using a perfect-aggregation (PA) FEEL simulation, where the per-round local model update fragments from each active device and BS are saved (see Section~\ref{sec:dataset_collection}). We then pre-train the communication stack end-to-end on this dataset, yielding a jointly learned URA codebook and AMP-DA-Net decoder, with the option to also include the quantisation stage within the learned scope. After pre-training, the communication model is fixed for deployment and reused across tasks and communication rounds without retraining, much like a conventional non-learned design. To clarify, this offline communication-layer training is separate from the FEEL global model learning task. Devices still perform local learning on their private datasets and transmit updates each round, while the learned encoder-decoder remains fixed and is only used to communicate and aggregate those updates over the uplink.

\subsection{Encoding}
At the $t$-th global round, after error feedback has been applied as in the communication model \eqref{s_update}, the $k$-th active device holds the update vector $\mathbf{s}_k^t \in \mathbb{R}^W$. This vector is split into $J$ fragments of length $d$, $\mathbf{u}_{k,i}^t \in \mathbb{R}^d$, as
\begin{equation}
    \mathbf{s}_k^t
    ~=~
    \big[ (\mathbf{u}_{k,1}^t)^\top, \mathbf{u}_{k,2}^t)^\top,\ldots, (\mathbf{u}_{k,J}^t)^\top \big]^\top,
\end{equation}
where index $i \in \{1,2,\ldots,J\}$ labels the fragment position within the update. To simplify notation, we drop the device and global round indices $k$ hereafter (e.g., refer to a single fragment as $\mathbf{u}_i$ instead of $\mathbf{u}_{k,i}^t$). Each fragment is then quantised using the current vector-quantisation codebook $\mathbf{Q} = [\mathbf{q}_1,\mathbf{q}_2,\ldots,\mathbf{q}_{n}] \in \mathbb{R}^{d\times n}$, with $n$ being the number of possible codewords, broadcast by the BS at the start of the round. Next, vector quantisation is performed via a nearest-neighbour search,
\begin{equation}
    \hat{\mathbf{q}}_i
    ~=~
    \arg\min_{\mathbf{q}\in\mathbf{Q}}
    \big\|\mathbf{u}_i - \mathbf{q}\big\|_2,
    \label{eq:vq_nn}
\end{equation}
and the index of $\hat{\mathbf{q}}_i$ within the codebook is used to select a column of the shared URA codebook $\mathbf{C}=[\mathbf{c}_1,\mathbf{c}_2,\ldots,\mathbf{c}_n] \in \mathbb{R}^{l\times n}$, with its columns, $\mathbf{c}_i$, as possible codewords. In each transmission slot (occupying $l$ channel uses), every active device transmits exactly one codeword $\mathbf{c}_j$ from this URA codebook, associated with its current fragment. A channel use may correspond to an OFDM symbol, a group of subcarriers used in parallel, or a time-slot, with the only requirement being that all devices use the same resources simultaneously in a given slot so that their signals superpose linearly at the BS. No user-specific preambles or explicit device identifiers are used, as all devices share the same codebooks.

\subsection{Sparse Recovery Formulation}\label{sec:sparse_recovery}
It is useful to view the resulting uplink communication task as a sparse recovery problem, as this motivates the use of the URA representation and the learned compressed-sensing decoder employed in our design. Consider a single fragment slot. Each active device selects one column of $\mathbf{C}$ (corresponding to its quantised message fragment) and transmits it in that slot. Let $\mathbf{x}\in\mathbb{R}^n$ denote the activity vector whose $i$-th entry counts how many devices selected codeword $\mathbf{c}_i$. The vector $\mathbf{x} \in \mathbb{N}_0^{n}$ is therefore non-negative, integer-valued, and sparse, with $\|\mathbf{x}\|_0 \ll n$ and $\sum_{i=1}^n x_i = K_a$. The received signal in a transmission slot, $\mathbf{y} \in \mathbb{R}^{l}$, can then be written as
\begin{equation}\label{eq:compressed_sensing}
  \mathbf{y} = \mathbf{C}\mathbf{x} + \mathbf{n},
\end{equation}
where $\mathbf{n}\in \mathbb{R}^{l}$ is the additive white Gaussian noise (AWGN) with IID entries with zero mean and variance $\sigma^2$; i.e., $\mathbf{n}\sim\mathcal{N}(0,\sigma^2 \mathbf{I}_l)$. The BS first recovers $\mathbf{x}$ from these noisy measurements, which is then used to recover the set of quantised messages. This is a canonical compressed-sensing problem, but with additional structure induced by the URA codebook and device activity constraints, and enables simultaneous transmissions without preambles or explicit device identifiers.

\subsection{Decoding}
The BS receives the noisy linear superposition of codewords in each fragment slot as in~\eqref{eq:compressed_sensing} and uses the learned decoder to obtain an estimate of the corresponding activity vector, $\hat{\mathbf{x}}$. The recovered indices and their estimated counts are then mapped back to entries of the quantisation codebook $\mathbf{Q}$, producing a set of reconstructed fragments $\{\hat{\mathbf{u}}_i\}$ across all devices. Recovered fragments are then passed to the symmetric aggregation rule in~\eqref{g} to obtain $\mathbf{g}^t$, followed by the global update rule \eqref{update}. The updated global model is then broadcast to devices along with the updated quantisation codebook. We next explain how the quantisation codebook is constructed and updated.

\subsection{Quantisation Codebook Construction}
\label{quant_codebook}
At the start of each round, the BS trains locally on its own dataset $\mathcal{D}_0$ and fragments its update into vectors of length $d$. These fragments are then clustered with $n$ centroids, to form an initial quantisation codebook $\mathbf{Q_{temp}}\in\mathbb{R}^{d\times n}$, whose columns $\{\mathbf{q}_1,\mathbf{q}_2,\ldots,\mathbf{q}_{n}\}$ correspond to centroids produced from clustering. The codebook is constructed using the \textit{k-means$++$} initialisation~\cite{arthur2007_kmeanspp}, which provides a well-spread, diversity-maximising set of centroids\footnote{Additional Lloyd iterations were tested but led to overfitting of the BS’s local distribution, degrading quantisation performance for other devices.}. The BS then quantises its own fragments using $\mathbf{Q_{temp}}$, records the number of assignments $N_i$ to each centroid $\mathbf{q}_i$, and forms a normalised popularity distribution as
\begin{equation}
    \hat{\pi}_i = \frac{N_i}{\sum_{j} N_j}.
\end{equation}
The centroids are then ordered from the most to least popular according to $\hat{\pi}_i$, forming the final quantisation codebook $\mathbf{Q}$, which is broadcast to all devices. The motivation for this step is that the BS’s popularity distribution provides a proxy for other devices’ codeword usage. By applying the same ordering across all devices, the expected distribution of codeword usage is standardised across rounds and learning tasks, presenting a more consistent input distribution to the URA codebook during training and influencing its learned structure. In addition, the decoder can learn that earlier codewords are more likely to be active and exploit this as prior information during recovery. Note that ordering by similarity in cosine distance was also tested, but was much less impactful than popularity ordering.

We further improved the quantisation codebook design by constructing $\mathbf{Q_{temp}}$ in a curvature-informed feature space, which employs the Hessian or the empirical Fisher matrix of the global loss function. However, computing the true Hessian or the empirical Fisher matrix would require repeated second-order information from all devices and access to their local data, which is infeasible in a FEEL setting. Instead, the BS local training can be used to form a lightweight diagonal proxy that approximates relative curvature per dimension. Let $\mu \in \mathbb{R}^d$ denote the per-dimension empirical mean, with components $\mu_i$, computed over the BS’s fragments. A variance-based sensitivity estimate is then formed as
\begin{align}
    W_i = \frac{1}{\sqrt{\sigma_i^2 + \varepsilon}},  
\end{align}
for $i = 1,2,\ldots,d$, where $\sigma_i^2 = \operatorname{Var}(u_i - \mu_i)$ and $\varepsilon > 0$ is a small stabilising constant. Defining $\mathbf{W} = \mathrm{diag}(W_1,\dots,W_d)$, the BS applies the diagonal transformation
\begin{equation}
    \tilde{\mathbf{u}} = \mathbf{W}(\mathbf{u}-\boldsymbol{\mu}),
\end{equation}
which expands directions with low empirical variance (higher sensitivity) and compresses directions with high variance. The k-means$++$ initialisation is then performed in this transformed space to obtain centroids $\tilde{\mathbf{q}}_j$, which are mapped back to the original space via
\begin{equation}
    \mathbf{q}_j = \mathbf{W}^{-1}\tilde{\mathbf{q}}_j + \boldsymbol{\mu}.
\end{equation}
This produces a codebook whose centroids are biased toward dimensions that encode higher local curvature at the BS, hypothesised to improve quantisation (see Appendix~\ref{app:hessian_quant}) without requiring full Hessian or Fisher computation. 

\begin{figure*}[t]
    \centering
    \includegraphics[width=\textwidth]{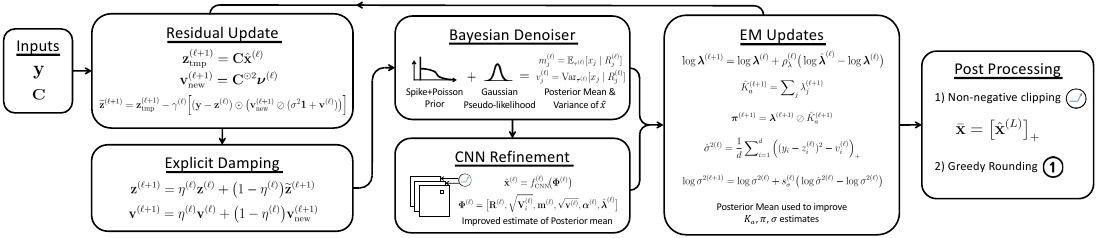}
    \vspace{-16pt}
    \caption{Decoder structure of the proposed AMP-DA-Net.}
    \vspace{-8pt}
    \label{fig:ampdanet}
\end{figure*}

\subsection{AMP-DA-Net: Proposed Learned Decoder} \label{amp_da_net}
The core component of the proposed solution is a learned unrolled decoder, AMP-DA-Net, which builds upon AMP, generalised AMP (GAMP)~\cite{rangan2011_generalized_amp}, AMP-Net~\cite{zhang2021_ampnet}, and the AMP-based digital aggregation (AMP-DA) decoder used in MD-AirComp~\cite{ qiao2024_massive_digital_ota_computation}. AMP frames decoding as a sequence of scalar denoising tasks by alternating residual updates with sparsity-promoting shrinkage and an Onsager correction term that keeps the effective noise approximately IID Gaussian. GAMP generalises this to arbitrary likelihoods and priors, splitting each iteration into an \emph{output update} in the measurement domain and an \emph{input update} in the codeword domain~\cite{rangan2011_generalized_amp}. AMP-Net then unrolls this iterative scheme into a fixed-depth network, replacing the shrinkage function with a convolutional neural network (CNN)-based denoiser and learning per-layer calibration parameters such as step size, scaling, and damping factors~\cite{zhang2021_ampnet}. 

Classical AMP-based decoders, including AMP-DA used in MD-AirComp, rely on analytically derived priors, fixed codebook constructions, and asymptotic assumptions (e.g., large random sensing matrices and well-matched noise statistics) to guarantee approximate Gaussianity of the residuals. In practical FEEL settings, however, these assumptions are systematically violated due to finite codebook sizes, structured URA designs, highly non-uniform codeword popularity, quantisation mismatch, and strong SNR variability across rounds. As a result, manually tuned AMP variants struggle to simultaneously maintain reliable sparse recovery, accurate activity estimation, and stable global convergence, particularly in the low-SNR regime. The proposed AMP-DA-Net addresses these limitations by learning from data how to compensate for these mismatches: the unrolled architecture jointly adapts residual scaling, damping, priors, and codebook geometry, thereby keeping the decoder stable and effective even when the classical AMP state-evolution assumptions no longer hold. This learned correction mechanism is the key reason the proposed approach achieves reliable convergence in regimes where analytically designed AMP-based schemes fail.

Next, we explain the working blocks of the proposed learned decoder, AMP-DA-Net, as depicted in Fig.~\ref{fig:ampdanet}. Since AMP-DA-Net processes each fragment slot independently and identically, regardless of the global round index, we explain the decoding process only for a single fragment slot.

\vspace{3pt}
\noindent
\textbf{Initialisation:} 
For every fragment, AMP-DA-Net runs for $L$ layers
At layer $\ell$, $\ell = 0,\dots,L-1$, it maintains $\hat{\mathbf{x}}^{(\ell)} \in \mathbb{R}^{n}$, $\boldsymbol{\nu}^{(\ell)} \in \mathbb{R}^{n}$, $\mathbf{z}^{(\ell)} \in \mathbb{R}^{l}$, and $\mathbf{v}^{(\ell)} \in \mathbb{R}^{l}$,
where $\hat{\mathbf{x}}^{(\ell)}$ is the current estimate of the activity-count vector in the \emph{codeword domain}, $\boldsymbol{\nu}^{(\ell)}$ is its per-codeword variance proxy, $\mathbf{z}^{(\ell)}$ is an Onsager-corrected estimate of noiseless measurement $\mathbf{C}\hat{\mathbf{x}}^{(\ell)}$ in the \emph{measurement domain}, and $\mathbf{v}^{(\ell)}$ is the associated per-measurement variance proxy. The corresponding residual is
\begin{equation}
\mathbf{r}^{(\ell)} = \mathbf{y} - \mathbf{z}^{(\ell)}.
\end{equation}
At the beginning of each FEEL round, and for every fragment slot, these variables are freshly initialised as $\hat{\mathbf{x}}^{(0)} = \mathbf{0}$, $\boldsymbol{\nu}^{(0)} = \mathbf{1}$, $\mathbf{z}^{(0)} = \mathbf{y}$, and $\mathbf{v}^{(0)} = \mathbf{1}$,
and are then updated layer by layer by the output and input update steps described next.
\vspace{3pt}

\noindent
\textbf{Output Block:}
Given the state $\big(\hat{\mathbf{x}}^{(\ell)}, \boldsymbol{\nu}^{(\ell)}, \mathbf{z}^{(\ell)}, \mathbf{v}^{(\ell)}\big)$ and a per-layer estimate of the noise variance $\sigma^2$ (see Section~\ref{em_updates}), the output block updates the measurement-domain mean and variance. First, it projects the codeword-domain estimate and variance through the codebook:
\begin{align}
\mathbf{z}_{\mathrm{tmp}}^{(\ell+1)} &= \mathbf{C}\,\hat{\mathbf{x}}^{(\ell)},\\
\mathbf{v}_{\mathrm{new}}^{(\ell+1)} &= \mathbf{C}^{\odot2}\,\boldsymbol{\nu}^{(\ell)}.
\end{align}
We also define the effective noise variance
\begin{equation}
\mathbf{d}^{(\ell)} = \sigma^2 \mathbf{1} + \mathbf{v}^{(\ell)}.
\end{equation}
In classical GAMP, the output step rescales this residual by $\mathbf{v}_{\mathrm{new}}^{(\ell+1)} \oslash \mathbf{d}^{(\ell)}$ and applies an Onsager correction. In the proposed AMP-DA-Net, the corresponding gain and Onsager term are absorbed into a single learnable scalar $\gamma^{(\ell)}$, and we additionally apply a learnable damping factor $0 < \eta^{(\ell)} < 1$ to stabilise updates:
\begin{align}
\widetilde{\mathbf{z}}^{(\ell+1)} 
&= \mathbf{z}_{\mathrm{tmp}}^{(\ell+1)}
  - \gamma^{(\ell)} 
    \Big[\, \mathbf{r}^{(\ell)} \odot 
           \big(\mathbf{v}_{\mathrm{new}}^{(\ell+1)} \oslash \mathbf{d}^{(\ell)}\big) \,\Big],\\
\mathbf{z}^{(\ell+1)} 
&= \eta^{(\ell)} \mathbf{z}^{(\ell)} 
  + \big(1 - \eta^{(\ell)}\big)\,\widetilde{\mathbf{z}}^{(\ell+1)},\\
\mathbf{v}^{(\ell+1)} 
&= \eta^{(\ell)} \mathbf{v}^{(\ell)} 
  + \big(1 - \eta^{(\ell)}\big)\,\mathbf{v}_{\mathrm{new}}^{(\ell+1)}.
\end{align}
To constrain learning, $\gamma^{(\ell)}$ is parametrised via a centred $\tanh$ mapping and restricted to lie in $[0.3,2]$\footnote{Allowing values slightly above $1$ introduces controlled instability, shown to improve expressiveness and convergence~\cite{nareddy2025_intriguing_learnt_matrices}.}, while $\eta^{(\ell)}$ is obtained from a sigmoid gate and thus lies in $(0,1)$.
\vspace{3pt}

\noindent
\textbf{Input Block:} 
The input block performs codeword-domain denoising. After the measurement-domain update, each component of the current estimate $\hat{\mathbf{x}}^{(\ell)}$ is viewed as passing through a scalar pseudo-channel. We first form a per-measurement precision vector as
\begin{equation}
  \boldsymbol{\kappa}^{(\ell)}
  \;=\;
  \beta^{(\ell)} \odot \big(\mathbf{d}^{(\ell)}\big)^{-1},
\end{equation}
where $\beta^{(\ell)}$ is a learnable per-layer scaling factor bounded via a centred $\tanh$ mapping to $[0.5,2]$, and $\mathbf{d}^{(\ell)} = \sigma^2 \mathbf{1} + \mathbf{v}^{(\ell)}$ is the effective measurement variance from the output block. Intuitively, when $\sigma^2 + v_d^{(\ell)}$ is small, the corresponding precision $\kappa_d^{(\ell)}$ becomes large and the measurement at index $d$ is trusted more in the subsequent update. Projecting $\boldsymbol{\kappa}^{(\ell)}$ through the squared codebook yields a pseudo-channel precision and variance in the codeword domain:
\begin{align}
  \boldsymbol{\psi}^{(\ell)} 
  &\;=\; \mathbf{C}^{\odot 2\,\top} \boldsymbol{\kappa}^{(\ell)},\\  \mathbf{V}_i^{(\ell)} 
  &\;=\; \big(\boldsymbol{\psi}^{(\ell)}\big)^{-1},
\end{align}
and a matched-filter term
\begin{equation}
  \boldsymbol{\rho}^{(\ell)}
  \;=\;
  \mathbf{C}^\top\!\big(\boldsymbol{\kappa}^{(\ell)} \odot \mathbf{r}^{(\ell)}\big).
\end{equation}
Component-wise, this corresponds to
\begin{align}
  \psi_{j}^{(\ell)} &= \sum_{m} \kappa_m^{(\ell)} C_{mj}^2,\\
  \rho_{j}^{(\ell)} &= \sum_{m} \kappa_m^{(\ell)} C_{mj} r_m^{(\ell)},
\end{align}
which matches the standard GAMP input update for an AWGN output channel~\cite{rangan2011_generalized_amp}. The vector $\boldsymbol{\psi}^{(\ell)}$ thus plays the role of a pseudo-channel \emph{precision}, and its inverse $\mathbf{V}_i^{(\ell)}$ is the corresponding pseudo-channel \emph{variance} in the codeword domain. The pseudo-observation for each coefficient is then
\begin{equation}
  \mathbf{R}^{(\ell)}
  \;=\;
  \hat{\mathbf{x}}^{(\ell)} 
  \;+\; 
  \boldsymbol{\rho}^{(\ell)} \oslash \boldsymbol{\psi}^{(\ell)},
  \label{eq:ampdanet_pseudo_channel}
\end{equation}
with corresponding variance $\mathbf{V}_i^{(\ell)}$. Equivalently, each coordinate is treated as obeying a scalar Gaussian channel, which is then combined with a structured prior in the next step
\begin{equation}
  R_j^{(\ell)} = x_j + w_j, 
\end{equation}
where $w_j \sim \mathcal{N}\!\big(0, V_{i,j}^{(\ell)}\big)$. 

Each element $x_j$ is then de-noised under a spike-and-slab prior, where the spike is a point mass at zero, and the slab is a Poisson distribution, as
\begin{equation}
  p(x_j) = (1-\alpha_j)\,\delta_0 + \alpha_j \,\mathrm{Pois}(\lambda_j),
\end{equation}
with activity probability $\alpha_j$ and Poisson rate $\lambda_j$ for codeword $j$. This reflects that most codewords are unused (spike at zero), while active ones can be reasonably approximated by a Poisson-distributed count with mean $\lambda_j$. Combining this prior with the scalar pseudo-channel likelihood defined by $\mathbf{R}^{(\ell)}$ and $\mathbf{V}_i^{(\ell)}$ yields a temperature-smoothed posterior mean and variance
\begin{align}
  m^{(\ell)}_j &= \mathbb{E}_{\tau^{(\ell)}}[x_j \mid R^{(\ell)}_j],\\  v^{(\ell)}_j &= \mathrm{Var}_{\tau^{(\ell)}}[x_j \mid R^{(\ell)}_j],
\end{align}
where $\tau^{(\ell)} > 0$ is a learnable per-layer temperature (small $\tau^{(\ell)}$ is MAP-like, while larger $\tau^{(\ell)}$ smooths the posterior and improves stability at low SNRs). Collecting into vectors $\mathbf{m}^{(\ell)}$ and $\mathbf{v}^{(\ell)}$, the uncertainty state follows the Bayesian update
\begin{equation}
  \boldsymbol{\nu}^{(\ell+1)} = \mathbf{v}^{(\ell)}.
\end{equation}
Next, we augment the Bayesian moments with a small learned CNN denoiser that takes in a compact 6-channel codeword-domain feature map, including a standardised log-intensity input. This eliminates the global scale, enabling the CNN to focus on relative popularity patterns rather than absolute values. The 1D CNN $f_{\mathrm{CNN}}^{(\ell)}$ then produces a refined estimate
\begin{equation}
  \tilde{\mathbf{x}}^{(\ell)} 
  \;=\; f_{\mathrm{CNN}}^{(\ell)}\!\big(\boldsymbol{\Phi}^{(\ell)}\big),
\end{equation}
where the inputs include
\begin{equation}
\boldsymbol{\Phi}^{(\ell)}=
  \big[
    \mathbf{R}^{(\ell)},\,
    \sqrt{\mathbf{V}_i^{(\ell)}},\,
    \mathbf{m}^{(\ell)},\,
    \sqrt{\mathbf{v}^{(\ell)}},\,
    \boldsymbol{\alpha}^{(\ell)},\,
    \tilde{\boldsymbol{\lambda}}^{(\ell)}
  \big],
\end{equation}
with $\boldsymbol{\alpha}^{(\ell)}$ being the current activity parameter, and
\begin{equation} 
  \tilde{\boldsymbol{\lambda}}^{(\ell)} 
  \;=\; 
  \frac{\log \boldsymbol{\lambda}^{(\ell)} 
        - \mathrm{mean}_j\big(\log \boldsymbol{\lambda}^{(\ell)}\big)}
       {\mathrm{std}_j\big(\log \boldsymbol{\lambda}^{(\ell)}\big) + \varepsilon},
\end{equation}
where $\boldsymbol{\lambda}^{(\ell)}$ is the current rate. Finally, $\tilde{\mathbf{x}}^{(\ell)}$ is blended with the Bayesian mean using a learnable gate $0 \le \zeta^{(\ell)} \le 1$, as 
\begin{equation}
  \hat{\mathbf{x}}^{(\ell+1)} 
  \;=\; \big(1-\zeta^{(\ell)}\big)\,\mathbf{m}^{(\ell)} 
         \;+\; \zeta^{(\ell)}\,\tilde{\mathbf{x}}^{(\ell)}.
\end{equation}

\noindent
\textbf{Expectation Maximisation (EM) Updates:} \phantomsection\label{em_updates}
To adapt across different setups, several latent parameters are refined at each layer using damped EM-style updates that stabilise learning while preserving the structure of the Poisson-spike prior. From the discrete posterior table in the input block, the batch-averaged posterior means provide a new proposal $\hat{\boldsymbol{\lambda}}^{(\ell)}$ for the Poisson rates associated with each codeword index. Rather than replacing the current values, the update is performed in the log-domain using a confidence-controlled interpolation:
\begin{equation}
    \log \boldsymbol{\lambda}^{(\ell+1)}
    \;=\; \log \boldsymbol{\lambda}^{(\ell)}
    \;+\; \rho_{\lambda}^{(\ell)}
          \big(\log \hat{\boldsymbol{\lambda}}^{(\ell)} 
               - \log \boldsymbol{\lambda}^{(\ell)}\big),
\end{equation}
where $0 < \rho_{\lambda}^{(\ell)} < 1$ is a scalar step size set by a posterior confidence statistic. This is calculated as the batch-averaged ratio $\overline{m}_j^{\,2}/\overline{v}_j$ (squared posterior means over posterior variances) across codewords, so that more confident posteriors trigger larger EM steps. The reason for the log-domain interpolation is that the Poisson rate is a positive, scale-dependent intensity measurement, so relative (multiplicative) changes carry consistent meaning, whereas additive changes do not, e.g, doubling activity has a consistent meaning, while adding a fixed offset does not. Once updated, the total activity and popularity distribution follow naturally as
\begin{align}
    \hat{K}_a^{(\ell+1)} &= \sum_j \lambda_j^{(\ell+1)}, \\    \boldsymbol{\pi}^{(\ell+1)} &= 
      \frac{\boldsymbol{\lambda}^{(\ell+1)}}{\hat{K}_a^{(\ell+1)}}.
\end{align}
The Bernoulli activity prior (spike mass) $\boldsymbol{\alpha}^{(\ell)}$ is tied to the updated rates through $\alpha_j^{(\ell+1)} \approx 1 - e^{-\lambda_j^{(\ell+1)}}$ and further stabilised via a learned per-layer mixing gate that blends this prior with the posterior activity estimates.

The noise variance $\sigma^2$ is refined by an EM-inspired moment-matching step on the residual energy, after which the proposed value is merged with the current one in the log-domain through a learnable gate $s_\sigma^{(\ell)}\!\in(0,1)$, following the implementation of~\cite{qiao2024_massive_digital_ota_computation}. These damped EM-style refinements enable AMP-DA-Net to generalise across SNRs, tasks, and rounds, while maintaining numerical stability. 

A keen reader may note that, in the underlying FEEL model, the true number of active devices, $K_a$, is shared across all fragments within a round; however, in our implementation, the EM updates above are applied independently to each fragment during decoding. The reason is that enforcing an explicit equality constraint on $\hat{K}_a$ across all fragments within the unrolled network would require cross-fragment coupling and substantially higher memory usage per mini-batch. This proved challenging to implement and so is left for future work. This effect is also partly mitigated by averaging the fragments' $\hat{K}_{a,i}$ at the end of the round and using the aggregate in the global model update. 
\vspace{3pt}

\noindent
\textbf{Post-processing:} \phantomsection\label{post_process}
A final correction step is applied to enforce the known structure of the activity vector. The decoder output is first clipped to be non-negative, and then projected onto the set of non-negative integer vectors whose entries sum to the estimated activity level $\hat{K}_a$. This projection is performed by a greedy rounding procedure that approximately minimises $\|\hat{\mathbf{x}} - \mathbf{x}\|_2$ subject to $\mathbf{x}\!\ge\!0$ and $\sum_j x_j = \hat{K}_a$. The resulting integer count vector is then de-quantised to recover the transmitted fragments, which are aggregated to produce the global update.

\vspace{3pt}
\noindent
\textbf{Dataset Collection:} \phantomsection\label{sec:dataset_collection}
Encoder-decoder training data is generated using a PA FEEL pipeline with error feedback and varying activity levels $K_a$. We use the same image-classification FEEL task, activity range $K_a$ and heterogeneity level as in the experimental setup in Section~\ref{sec:results}, and a ResNet with an initial 3×3 convolution (16 channels), followed by three residual stages of three basic blocks each at widths {16,32,64} with stride-2 downsampling between stages, global average pooling, and a final linear classifier. From this simulation, the raw model-update fragments of active devices are collected, along with their corresponding BS local model updates. This allows us to perform the quantisation codebook creation and sparse encoding stages during pre-training, as well as unrolled decoder learning, enabling end-to-end training with optional quantisation awareness. Note that the training dataset can be enhanced by combining data from multiple simulations of different setups to increase the variability of scenarios the learned communication layer sees during training, but this was not found to be necessary (see Section~\ref{sec:general}). The resulting dataset captures realistic FEEL trends such as non-uniform codeword usage and per-device biases.
\vspace{3pt}

\noindent
\textbf{Codebook Representation:} \phantomsection\label{sec:codebook_rep}The URA codebook at the encoder is jointly trained with the decoder using a two-matrix parameterisation $\mathbf{C}_\text{syn} = \mathbf{DW}$, where $\mathbf{D} \in \mathbb{R}^{n\times l}$ stores base codeword vectors and $\mathbf{W} \in \mathbb{R}^{l\times l}$ is a learned shear/rotation\footnote{Initial results showed an accuracy improvement of about $1.5\%$, compared to single matrix representation.}. Following the SimVQ parameterisation, this improves gradient flow, avoids \textit{dead codewords} (rarely used ones), and stabilises training compared to learning $\mathbf{C}$ directly \cite{SimVQ}. Rows of $\mathbf{C}_\text{syn}$ are renormalised to unit $\ell_2$ norm after each update to maintain equal codeword power. For the quantisation codebook $\mathbf{D}$, Gaussian and Bernoulli initialisations were tested, with the former proving most effective. $\mathbf{W}$ was initialised as identity. This parameterisation provided smoother training and slightly higher recovery accuracy.
\vspace{3pt}

\noindent
\textbf{Loss Function:} \phantomsection\label{sec:loss_func}
The training objective combines several components: (i) a MSE reconstruction term $\|\hat{\mathbf{x}}_i - \mathbf{x}_i\|_2^2$, (ii) an $\ell_1$ sparsity penalty normalised by the ground-truth scale, (iii) an orthogonality regulariser $\|\mathbf{W}^{\top}\mathbf{W}-\mathbf{I}\|_F^2$ to keep the SimVQ transform well-conditioned, and (iv) an active-device estimation MSE $(\hat K_{a,i}-K_{a,i})^2$. We also experimented with a quantisation-error term that penalised the normalised-squared residual between each device’s pre- and post-quantisation updates, encouraging the learned decoder and codebook to jointly reduce end-to-end quantisation distortion (see Section~\ref{sec:variants_sol}).
\vspace{3pt}

\noindent
\textbf{Hyper-parameters:} \phantomsection\label{sec:hyperparameter} Pre-training used $64{,}000$ samples for training, $8{,}000$ for validation, and $10{,}000$ held-out samples for testing. Batch size was $64$, with a maximum of $500$ epochs and early stopping (patience $20$, tolerance $10^{-6}$). Optimisation used Adam with a learning rate of $10^{-4}$, halved if the validation loss failed to improve for $10$ epochs. Gradients were accumulated over per-round blocks, with one optimiser update applied per block. The decoder employed $10$ unrolled layers, each with a 1D CNN denoiser (32 filters, kernel size $3$), and the Bayesian-CNN mixing gate was initialised to $0.85$. Loss weights were set to $\lambda_{1}=0.01$, $\lambda_{W}=0.001$, and $\lambda_{K}=0.01$. For experiments including the optional quantisation-error term, $\lambda_{q}={0.1}$ was tested.
\vspace{3pt}

\noindent
\textbf{Computational complexity:} The computational complexity of AMP-DA-Net scales linearly with the number of unrolled layers and URA codebook size, and is dominated by parallelisable matrix–vector operations, as in classical AMP-based decoders. The additional 1D CNN refinement applied per layer scales linearly with codebook size, approximately quadratically with the number of filters, and linearly with the kernel width, and these architectural parameters are tunable design choices. Training is performed offline, while online inference uses a fixed number of layers, providing consistent per-round latency and making the approach compatible with practical BS processing budgets.

\section{Results}
\label{sec:results}
The proposed setup was evaluated against the AMP-DA algorithm (the state-of-the-art) from different aspects\footnote{All codes are available at \url{https://github.com/tonytarizzo/AMP-DA-Net}}\textsuperscript{,}\footnote{All simulations are done using Imperial College London's high-performance computing (HPC) facility \cite{Imperial_HPC}}. Each device holds a local subset of CIFAR-10, with the global dataset split into $20\%$ IID and $80\%$ non-IID data. $10{,}000$ samples were randomly assigned to all devices, while the remaining $40{,}000$ were label-sorted into contiguous shards and then distributed sequentially to induce heterogeneity. The number of active devices $K_a$ was drawn uniformly from $[7,13]$, with total devices set to $K_t=40$. The server aggregated updates using FedAvg\footnote{Although we use FedAvg here, other federated optimizers such as FedAvgM, FedAdam, and related adaptive methods can be incorporated without modification to the surrounding setup~\cite{qiao2024_massive_digital_ota_computation, reddi2021adaptive_federated_optimization}.}. We set the fragment dimension to $d=20$ and used a URA codeword length of $l=64$ with a total of $n=128$ codewords.

\begin{figure}[t]
    \vspace{-10pt}
    \centering
    \includegraphics[width=.9\columnwidth]{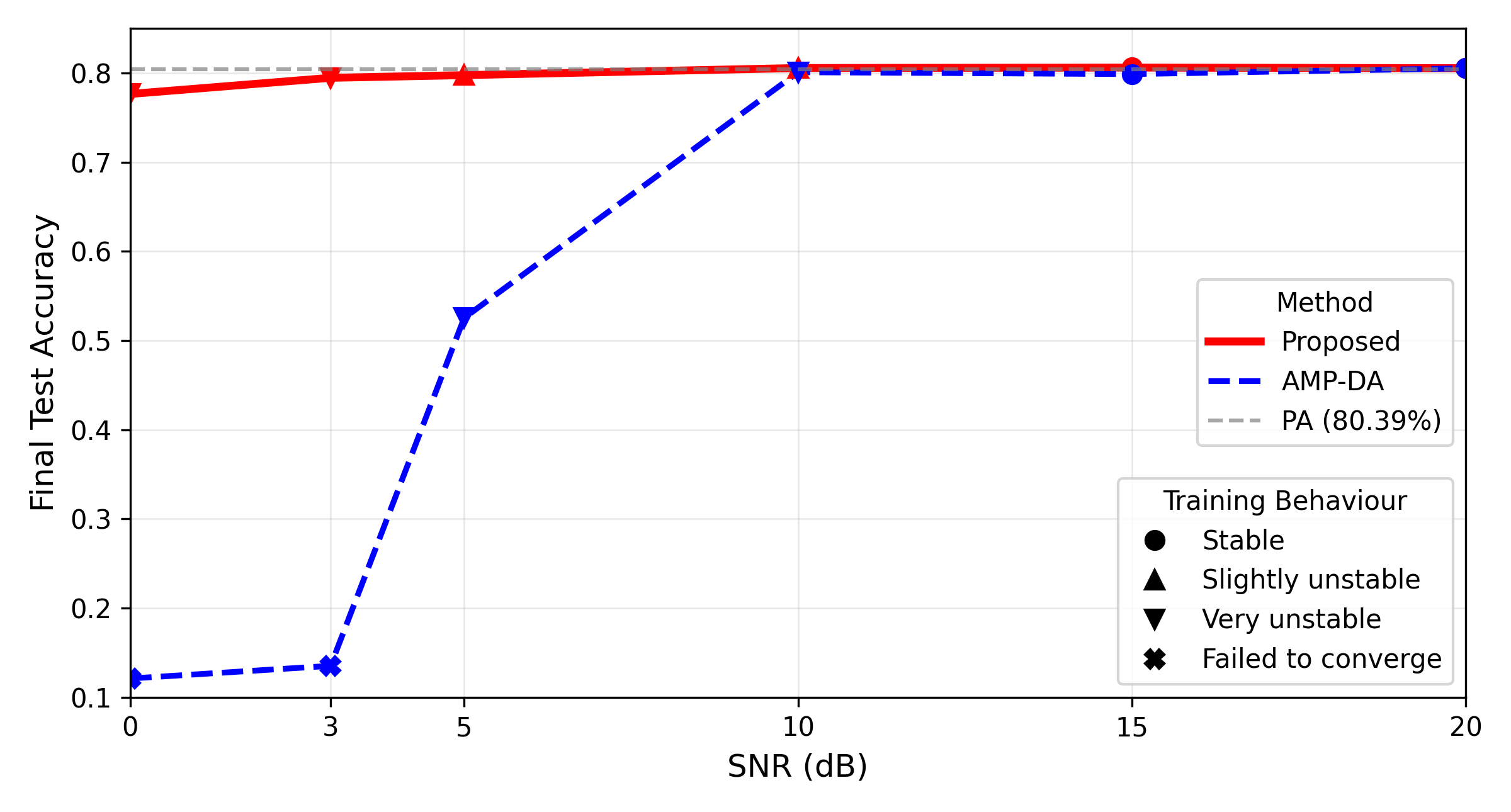}
    \vspace{-10pt}
    \caption{Final test accuracies and training stability.}
    \label{fig:fed_acc}
\end{figure}

\begin{table}[t]
\centering
\caption{Test accuracies across SNR (dB) (PA accuracy: 0.804).}
\label{tab:fed_acc}
\vspace{2pt}
\begin{tabular}{lcccccc}
\toprule
Method $\setminus$ SNR & 0 & 3 & 5 & 10 & 15 & 20 \\
\midrule
AMP-DA & 0.139 & 0.135 & 0.525 & 0.801 & 0.799 & \textbf{0.805}\\
Proposed & \textbf{0.776} & \textbf{0.795} & \textbf{0.798} & \textbf{0.806} & \textbf{0.806} & \textbf{0.805} \\
\bottomrule
\end{tabular}
\end{table}

As shown in Table~\ref{tab:fed_acc} and Fig.~\ref{fig:fed_acc}, the proposed method consistently outperforms AMP-DA across the entire SNR range, approaching the PA benchmark at low SNRs and surpassing it at higher SNRs. This can be attributed to slight variations arising from noise and imprecise decoding, which can be exploited during convergence. The proposed method improves reliability at higher SNRs, but more strikingly, it provides much stronger recovery and substantially more stable convergence at lower SNRs, thereby extending the operable range by more than $10$ dB. Note that we define SNR as the ratio of total received signal power to noise power, consistent with MD-AirComp's definition. Hence, for a fixed SNR, the noise level scales linearly with the number of active devices.

\begin{figure}[t]
    \vspace{-10pt}
    \centering
    \includegraphics[width=.9\columnwidth]{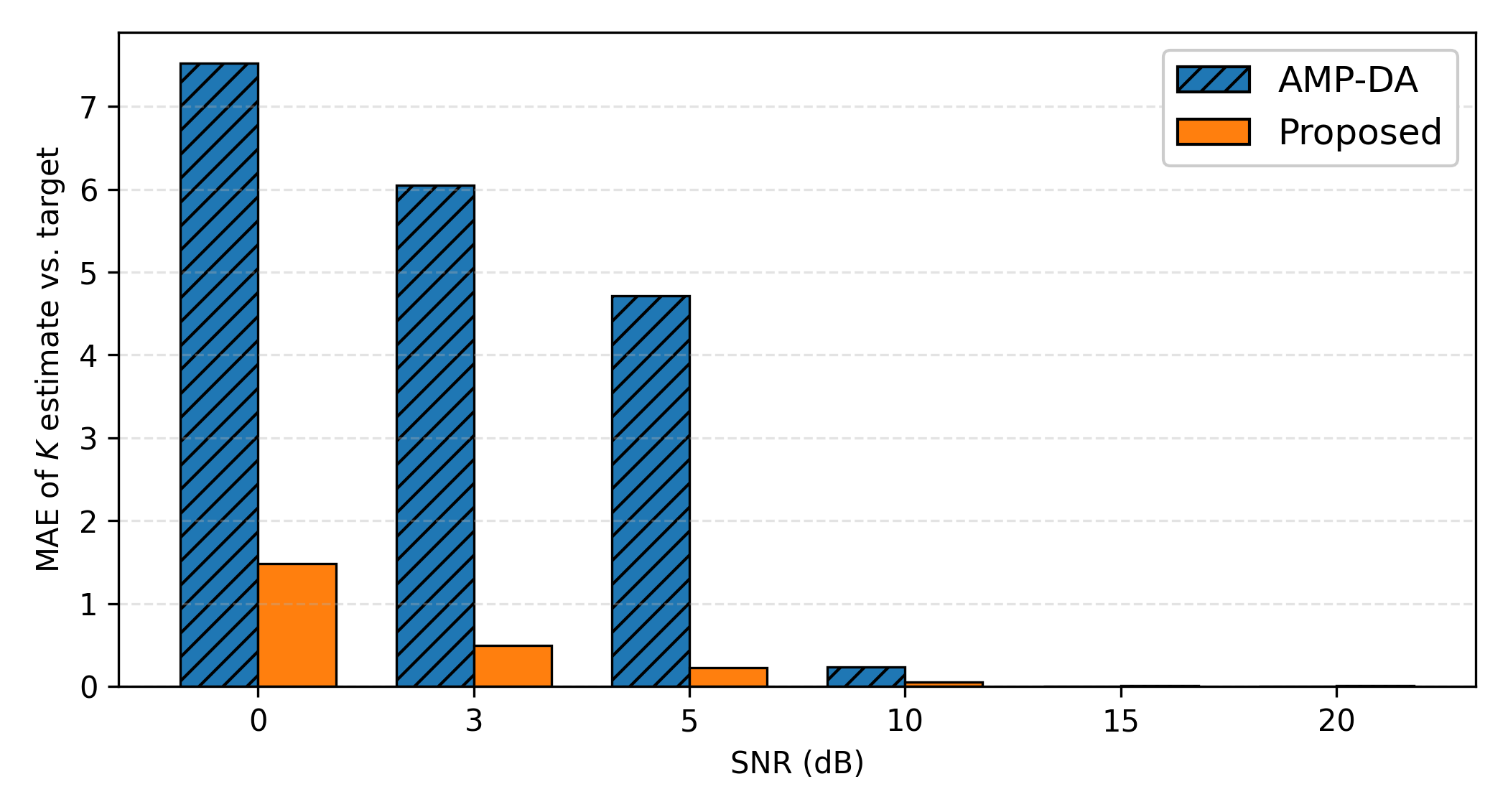}
    \vspace{-10pt}
    \caption{Mean average error (MAE) for $\hat K_a$ (unrounded).}
    \label{fig:mae}
\end{figure}

As shown in Fig.~\ref{fig:mae}, the accuracy of active-device estimation is also greatly improved, as measured by the mean absolute error (MAE)
$\mathrm{MAE} 
= \frac{1}{N_{\mathrm{round}}} \sum_{t=1}^{N_{\mathrm{round}}} 
   \bigl| K_a^{t} - \hat{K}_a^{t} \bigr|$,
where $K_a^{t}$ is the true number of active devices and $\hat{K}_a^{t}$ is the continuous (pre-rounding) estimate produced by the decoder in round $t$. Final estimates remain well within the $\pm0.5$ margin required for stable scaling in \eqref{eq:erm_standard} until very low SNRs. This is crucial for the arithmetic mean aggregation method since underestimating $K_a$ amplifies the global update, causing instability, whereas overestimating $K_a$ attenuates updates and slows convergence. 

\begin{figure}[t]
    \vspace{-10pt}
    \centering
    \includegraphics[width=.9\columnwidth]{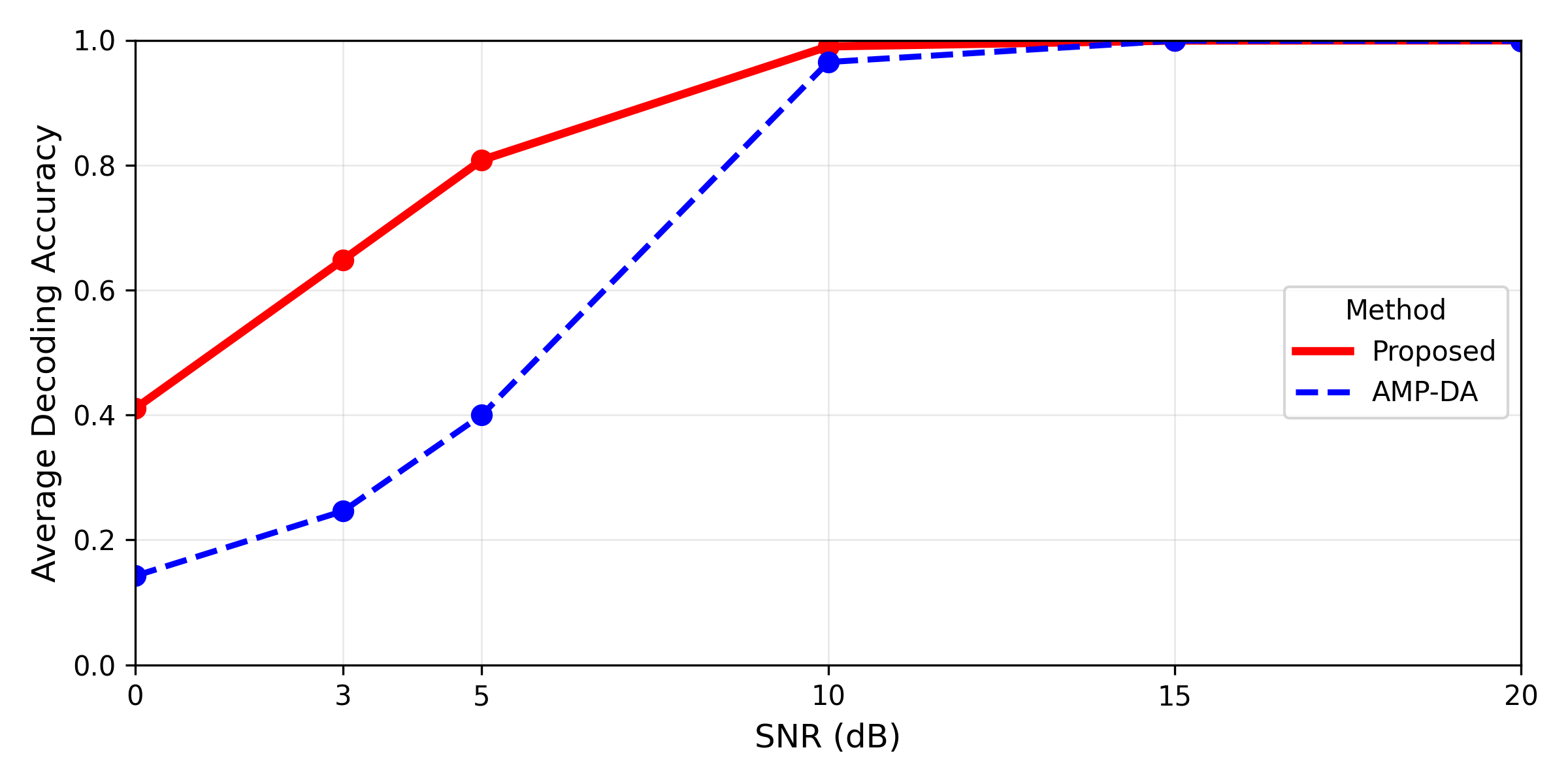}
    \vspace{-10pt}
    \caption{Decoding accuracy during inference.}
    \label{fig:decoding_acc}
\end{figure}

Finally, as observed in Fig.~\ref{fig:decoding_acc}, sparse vector recovery accuracy is consistently higher across all SNRs, highlighting the benefits of end-to-end learning of the digital codebook and the decoder. Accuracy was measured with a normalised $\ell_1$ accuracy score
$\mathrm{Acc} 
= \frac{1}{N_{\mathrm{test}}} \sum_{i=1}^{N_{\mathrm{test}}}
  \left(
    1 - \frac{\|\mathbf{x}_i - \hat{\mathbf{x}}_i\|_1}
             {\|\mathbf{x}_i\|_1}
  \right)$,
where $\mathbf{x}_i$ and $\hat{\mathbf{x}}_i$ are the true and recovered count vectors for test sample $i$. It is worth noting that some amount of inaccurate decoding is permissible in many OTA methods, as convergence can still be achieved even with reduced stability of global updates. This was observed in both the proposed method and AMP-DA.

\section{Analysis}
\begin{figure}[t]
    \vspace{-10pt}
    \centering
    \includegraphics[width=.9\columnwidth]{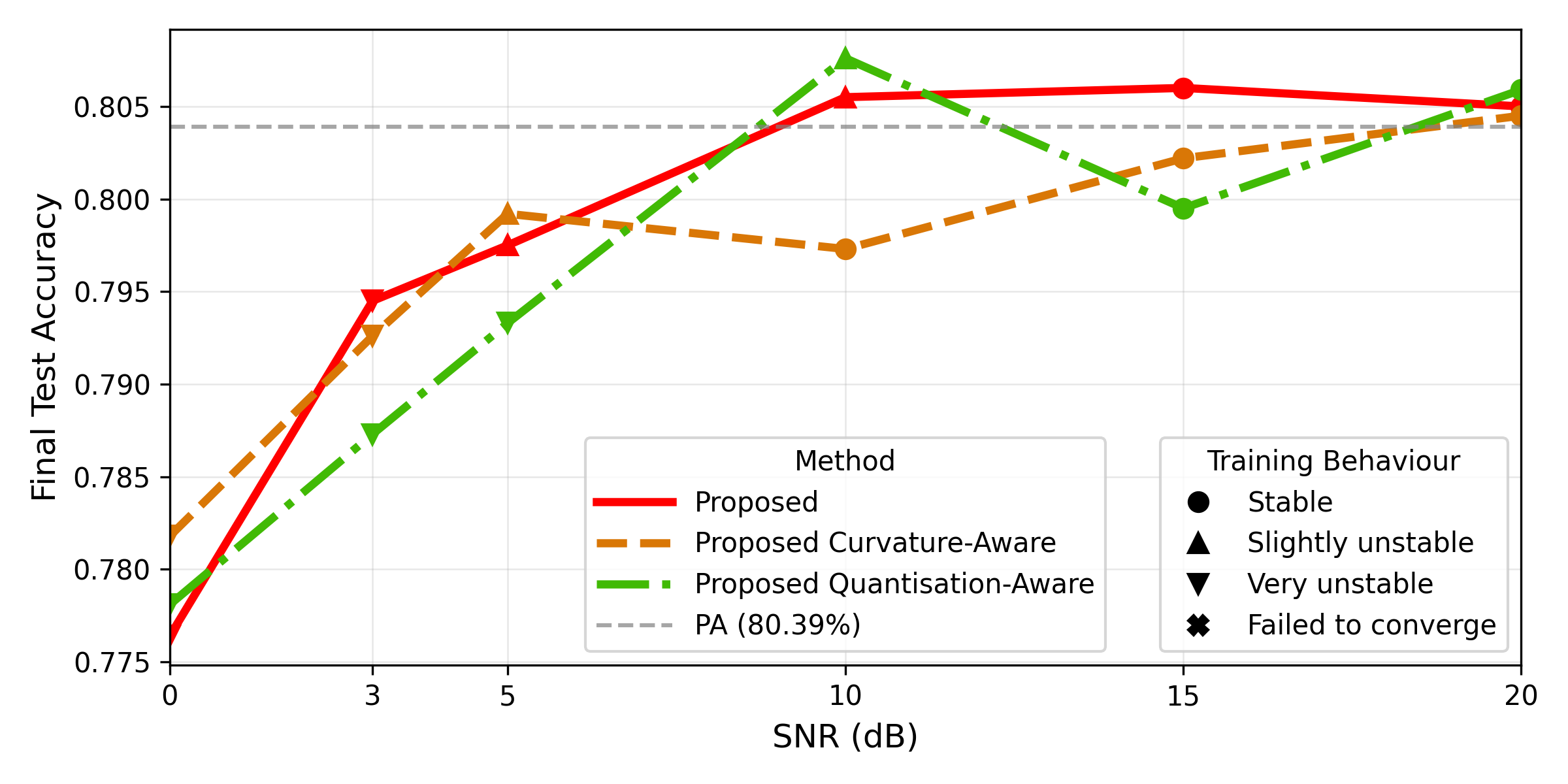}
    \vspace{-10pt}
    \caption{Curvature-aware and quantisation loss-aware variants.}
    \label{fig:fed_acc_2}
\end{figure}
\subsection{Variants of Proposed Method}\label{sec:variants_sol}
Alongside the main proposed method, we evaluated a curvature-aware variant that uses a statistical Hessian proxy during quantisation codebook construction (see Section~\ref{app:hessian_quant}), and a variant trained with an additional small quantisation-loss term calculated as the normalised-squared residual between pre- and post-quantisation messages (see Section~\ref{sec:loss_func}). As shown in Fig.~\ref{fig:fed_acc_2}, all three approaches remain viable, and their accuracy curves largely overlap across SNRs. The curvature-aware variant provides a slight improvement in stability as SNR decreases for a negligible increase in computational cost. Conversely, adding the quantisation-loss term consistently reduced stability and offered no compensating benefit in accuracy. Since none of the variants produce significant overall performance improvements, the baseline version is preferred.

\subsection{Generalisation} \label{sec:general}
\begin{figure}[t]
    \vspace{-10pt}
    \centering
    \includegraphics[width=.9\columnwidth]{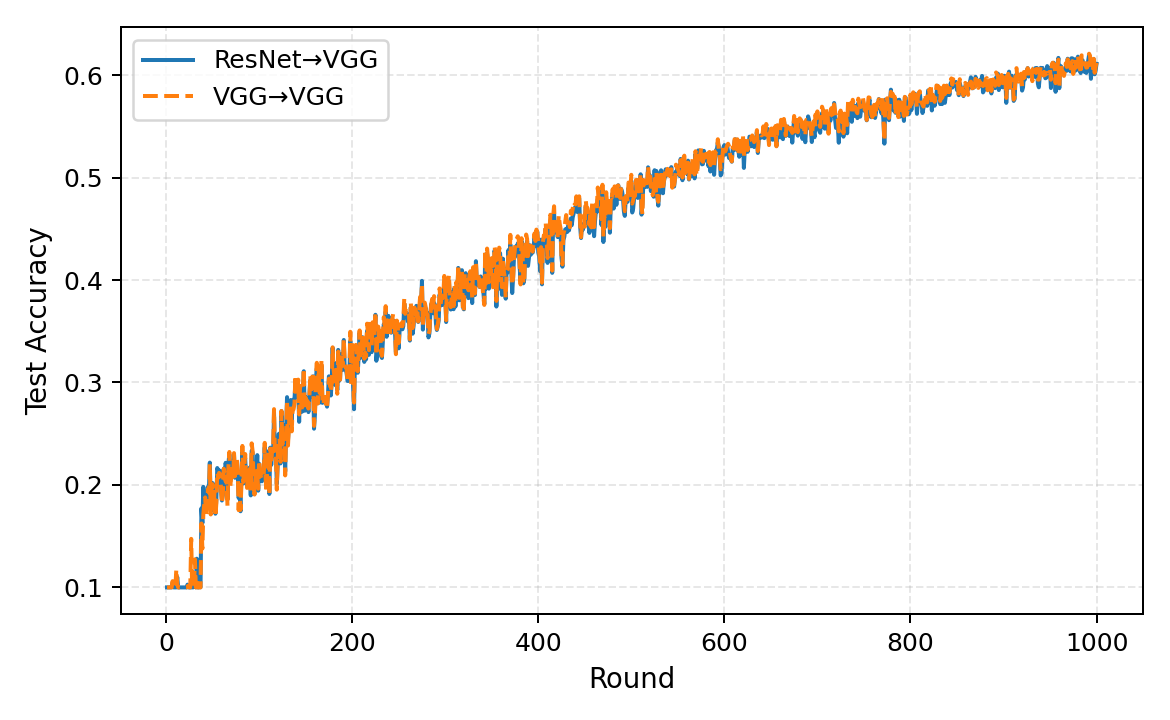}
    \vspace{-10pt}
    \caption{Communication models trained on ResNet or VGG, inference on VGG.}
        \vspace{-10pt}
    \label{fig:vgg_test}
\end{figure}

To evaluate generalisation directly, we trained the communication model using the ResNet global model ($269{,}722$ parameters) previously described in Section~\ref{sec:dataset_collection} to generate the dataset and perform pre-training. At inference time, however, the global model was replaced with a smaller \textit{visual geometry group} (VGG)-style network ($287{,}626$ parameters), and experiments were run at 10 dB SNR~\cite{vgg}. Specifically, the VGG-style CNN comprises three convolutional blocks (each with two 3×3 convolutions), max-pooling between blocks, global average pooling, and a linear classifier. As shown in Fig.~\ref{fig:vgg_test}, convergence remains essentially unchanged, with no noticeable degradation in performance. This demonstrates that a communication model learned using one architecture can transfer effectively to another, even when the downstream learning dynamics differ. This strong generalisation can be attributed to the communication design itself. The quantisation and sparse-coding stages reduce the communication problem to predicting discrete indices, while the learned codebook ordering ensures predictable codeword usage patterns across architectures. These choices simplify the decoder’s task, making unseen FL setups closely resemble those encountered during training and enabling strong out-of-distribution generalisation.

\subsection{Nonlinear Aggregation Methods}
To motivate and demonstrate the proposed setup's ability to work with aggregation rules beyond the arithmetic mean, we constructed a scenario in which $20\%$ of devices in each round (randomly selected and excluding the BS’s local training) transmitted corrupted updates. Specifically, the local model update was replaced with noise, but the device still transmitted the quantised representation of that corrupted signal. This models practical situations where a device undergoes local failure (e.g., due to hardware impairments), or adversarial interference, but still participates in uplink transmission, injecting structured bias into the aggregation process.

\begin{figure}[t]
    \centering
    \includegraphics[width=.9\columnwidth]{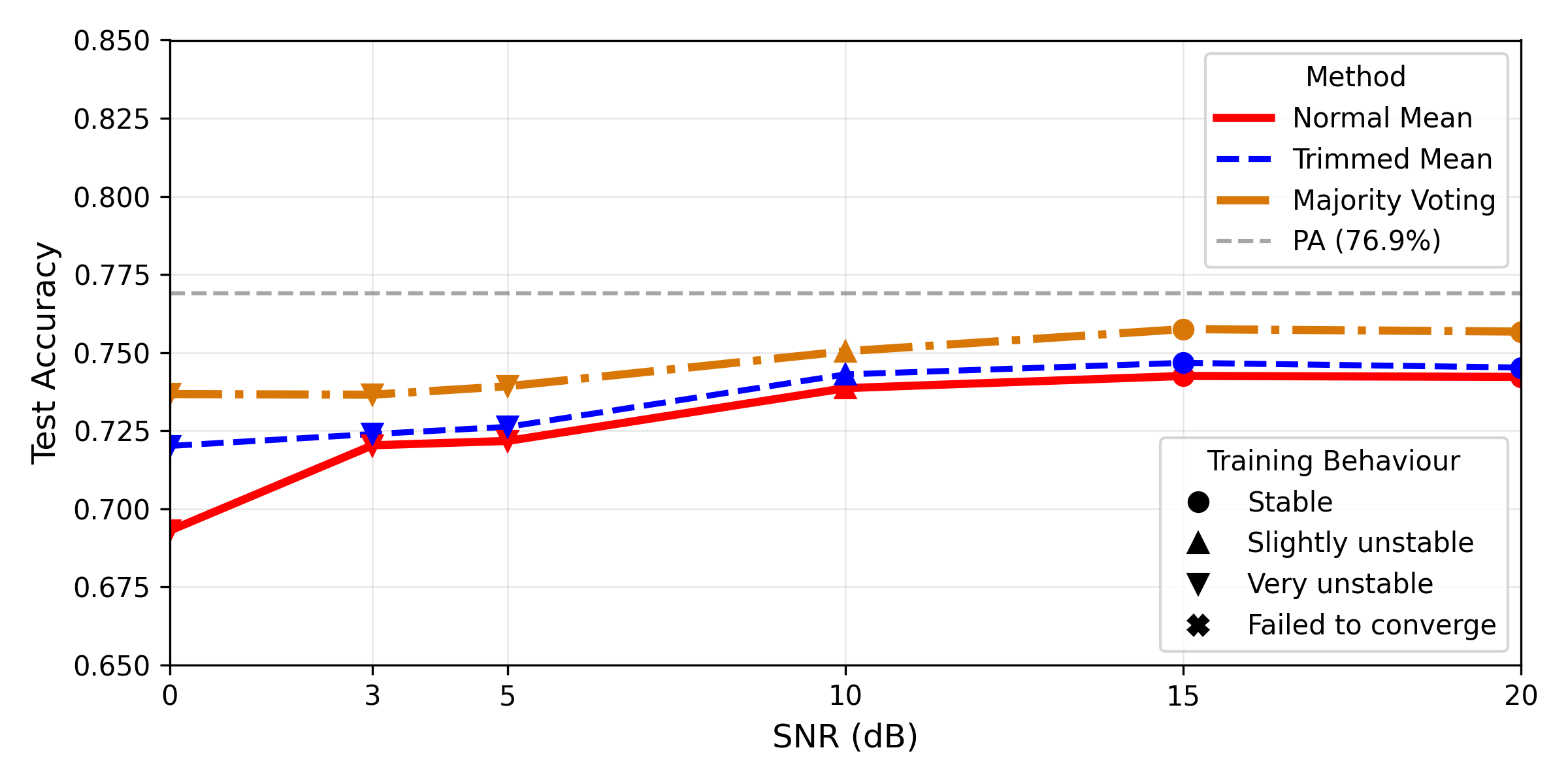}
    \vspace{-10pt}
    \caption{Inference results for mean, trimmed mean \& majority voting.}
    \label{fig:aggregate_comp}
    \vspace{-10pt}
\end{figure}

Because this corruption systematically distorts the arithmetic mean, we evaluated two nonlinear aggregation rules: trimmed mean (retaining $80\%$ of values) and majority voting (details for both in Appendix~\ref{app:nonlinear_aggregation})\footnote{Both rules operate directly on the recovered count vectors and require no changes to the learned communication layer, making them drop-in, robust alternatives to mean aggregation in our digital OTA pipeline.}. These were compared against the arithmetic mean, with all experiments run at an $\mathrm{SNR}=10\mathrm{dB}$. As shown in Fig.~\ref{fig:aggregate_comp}, both nonlinear methods substantially mitigate the bias introduced by corrupted devices, demonstrating that the learned communication channel is compatible with and robust to alternative aggregation mechanisms. 

During these corruption experiments, we found that error feedback had to be disabled to maintain stability. The underlying reason is that error feedback accumulates past residuals and adds them to each device’s update, causing device messages to drift in increasingly different directions over rounds. As these directions diverge, devices increasingly disagree on the appropriate codeword, making the quantisation codebook created at the BS a poor representation of the true distribution of message magnitudes and orientations. If this mismatch becomes too large (e.g., the residual error exceeds five times the magnitude of the current update), the codebook's ability to represent the device's weight vectors deteriorates, and the overall system becomes unstable and diverges. This highlights an important design trade-off. That is, error feedback can improve optimisation under clean conditions, but can also reduce stability under more challenging conditions. Tuning this stability-performance trade-off by adjusting the use of error feedback, modifying quantisation, or introducing round-wise adaptivity is a promising direction for future work.

\subsection{Codebook Construction}

\begin{figure}[t]
    \centering
    \includegraphics[width=\columnwidth]{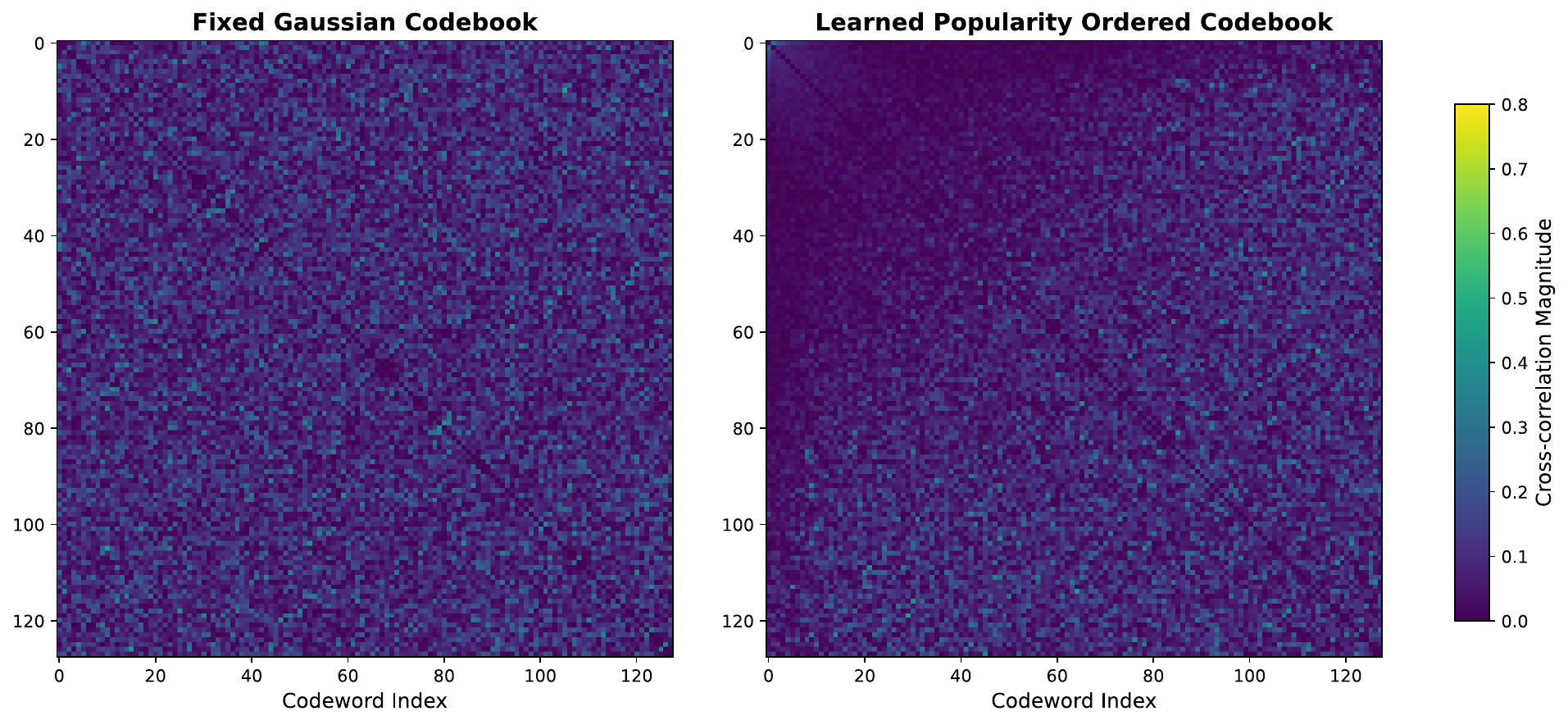}
    \caption{Popularity ordering \& learning effect on codeword cross-correlation.}
    \label{fig:heatmap}
\end{figure}

\begin{table}[t]
\centering
\caption{Codebook setup evaluation accuracy, SNR $=5\,\mathrm{dB}$.}
\label{tab:codebook_acc}
\begin{tabular}{lccc}
\toprule
Codebook & Initialisation & Ordering & Accuracy \\
\midrule
Learned & Gaussian   & Popularity & \textbf{0.932} \\
Learned & Gaussian   & None       & 0.859 \\
Fixed  & Gaussian    & None       & 0.696 \\
Fixed  & Bernoulli   & None       & 0.689 \\
\bottomrule
\end{tabular}
\vspace{-10pt}
\end{table}

The effects of codebook initialisation, learning, and ordering are summarised in Table~\ref{tab:codebook_acc}, demonstrating the benefits of learning the URA codebook and applying popularity-based ordering. Subsequent analysis of the URA codebook, as presented in Fig.~\ref{fig:heatmap}, showed that it learned to reduce pairwise cross-correlation more for popular codewords, while tolerating higher correlation for unpopular ones. Singular value analysis, which measures the spread of singular values, also showed a narrower range, indicating better conditioning than fixed baselines. As a result of popularity ordering improving decoding accuracy, it was found to significantly improve global learning stability and performance.

\section{Conclusion}
\label{sec:future_work}
We proposed a learned digital OTA communication framework for FEEL that improves robustness and performance in uplink-limited regimes without increasing communication overhead. By jointly learning the URA-based encoding, vector quantisation, and an unrolled AMP-style decoder, the proposed design adapts to practical violations of analytical assumptions and achieves reliable update recovery at significantly lower SNRs than existing digital OTA approaches. Experiments on highly non-IID datasets demonstrate substantial gains in low-SNR operation, leading to more stable and consistent global training, while generalising well across participation levels and model architectures.

Beyond mean aggregation, the proposed framework supports a broader class of symmetric aggregation rules by recovering the multiset of transmitted updates prior to aggregation. This enables robust alternatives, such as trimmed means and majority voting, within the same signalling and decoding pipeline, thereby improving resilience to corrupted device updates without redesigning the physical layer. Overall, the results highlight the benefit of integrating learning directly into the digital OTA communication stage, rather than limiting learning to higher-layer components. Future work can extend the framework to fading and multi-antenna channels and explore end-to-end designs that jointly learn precoding, OTA communication, and post-decoding aggregation, as well as adaptation to more severe system mismatches.

\appendices
\section{Motivation for k-means++ Codebook Construction}
\label{app:kmeans_quant}
This appendix demonstrates why $k$-means++ clustering is chosen to construct the quantisation codebook. This is not presented as a formal proof, but rather as an intuition linking minimising quantisation distortion to minimising the difference between the ideal (non-quantised) and quantised model updates. One caveat of this demonstration is that it does not account for a mismatch between the BS local update distribution and per-device update distributions. In practice, this mismatch can cause overfitting of the codebook to the BS dataset, which is why we found the well-distributed initialisation of $k$-means++ to be more stable.

\subsection{Setup and Assumptions}
We consider a single communication round and suppress the round index for clarity. Let
$\bar{\mathbf{s}} \;=\; \sum_{k=1}^{K} a_k \mathbf{s}_k$
denote the ideal aggregated update, where $\mathbf{s}_k$ is the (error-feedback corrected) local update of device $k$ and $a_k$ are aggregation weights (e.g., $a_k = 1/K$ for FedAvg). Devices transmit quantised updates $\mathcal{Q}(\mathbf{s}_k)$, leading to the quantised aggregate
$\hat{\mathbf{s}} =v\bar{\mathbf{s}} + \mathbf{e}$,
where $\bar{\mathbf{s}} = \sum_{k=1}^{K} a_k \mathcal{Q}(\mathbf{s}_k)$ and $\mathbf{e} \;=\; \sum_{k=1}^{K} a_k \big(\mathcal{Q}(\mathbf{s}_k) - \mathbf{s}_k\big)$ is the aggregate quantisation error. We assume that the global objective $F(\mathbf{w})$ is $L$-smooth, i.e.,
\begin{equation}
F(\mathbf{y}) \;\le\; F(\mathbf{x}) + \nabla F(\mathbf{x})^\top(\mathbf{y}-\mathbf{x}) + \frac{L}{2}\|\mathbf{y}-\mathbf{x}\|_2^2.
\end{equation}
We further assume that $\mathbf{e}$ is conditionally unbiased and approximately uncorrelated with the ideal aggregate $\bar{\mathbf{s}}$, i.e.,
\begin{equation}
\mathbb{E}\!\left[\mathcal{Q}(\mathbf{s}_k)\mid \mathbf{s}_k\right] = \mathbf{s}_k
\quad\Longleftrightarrow\quad
\mathbb{E}\!\left[\mathbf{e}_k \mid \mathbf{s}_k\right]=\mathbf{0},
\end{equation}
where $\mathbf{e}_k \triangleq \mathcal{Q}(\mathbf{s}_k)-\mathbf{s}_k$. This implies
\begin{equation}
\mathbb{E}[\mathbf{e}] = \mathbf{0},
\qquad
\mathbb{E}\!\left[\bar{\mathbf{s}}^\top \mathbf{e}\right]=0,
\end{equation}
These assumptions are common in analyses of stochastic quantisation schemes (e.g., QSGD~\cite{alistarh2017qsgd} and TernGrad~\cite{wen2017terngrad}), and are motivated by viewing the quantisation error as (approximately) zero-mean noise whose direction is not systematically aligned with the update direction when averaged across devices and dimensions~\cite{alistarh2017qsgd,wen2017terngrad}.

\subsection{Ideal vs.\ Quantised Model Updates}
The ideal and quantised updates are respectively
\begin{equation}
\mathbf{w}^{+} = \mathbf{w} - \eta \bar{\mathbf{s}},
\qquad
\tilde{\mathbf{w}}^{+} = \mathbf{w} - \eta(\bar{\mathbf{s}} + \mathbf{e}).
\end{equation}
Applying the descent lemma with $\mathbf{x}=\mathbf{w}$ and $\mathbf{y}=\tilde{\mathbf{w}}^{+}$ gives
\begin{equation}
F(\tilde{\mathbf{w}}^{+})
\;\le\;
F(\mathbf{w})
- \eta \nabla F(\mathbf{w})^\top (\bar{\mathbf{s}}+\mathbf{e})
+ \frac{L\eta^2}{2}\|\bar{\mathbf{s}}+\mathbf{e}\|_2^2.
\end{equation}
Subtracting the corresponding bound for the ideal update gives us an upper bound on the excess loss due to quantisation,
\begin{equation}
\begin{aligned}
    \Delta F
&\triangleq
F(\tilde{\mathbf{w}}^{+}) - F(\mathbf{w}^{+})\\
&\le
- \eta \nabla F(\mathbf{w})^\top \mathbf{e}
+ \frac{L\eta^2}{2}\big(2\bar{\mathbf{s}}^\top\mathbf{e} + \|\mathbf{e}\|_2^2\big).
\end{aligned}
\end{equation}
Taking expectations and using the conditionally unbiased and approximately uncorrelated assumptions leaves us with
\begin{equation}
\mathbb{E}[\Delta F]
\;\le\;
\frac{L\eta^2}{2}\,\mathbb{E}\|\mathbf{e}\|_2^2.
\end{equation}

\subsection{Implication for Quantisation Design}
The above bound shows that, under a worst-case smoothness assumption, the expected degradation from quantisation is controlled by $\mathbb{E}\|\mathbf{e}\|_2^2$ (up to constants $L$ and $\eta$). Writing $\mathbf{e}=\sum_{k=1}^K a_k \mathbf{e}_k$, if we assume the per-device quantisation errors $\{\mathbf{e}_k\}$ are independent across devices and satisfy $\mathbb{E}[\mathbf{e}_k]=\mathbf{0}$, then the cross terms vanish and
\begin{equation}
\begin{aligned}
    \mathbb{E}[\|\mathbf{e}\|_2^2]
&=
\sum_{k=1}^{K} a_k^2\, \mathbb{E}[\|\mathbf{e}_k\|_2^2]\\
&=
\sum_{k=1}^{K} a_k^2\, \mathbb{E}[\|\mathcal{Q}(\mathbf{s}_k)-\mathbf{s}_k\|_2^2].
\end{aligned}
\end{equation}

For FedAvg with $a_k = 1/K$, this reduces to minimising the average per-device quantisation MSE, which matches the objective function of $k$-means and $k$-means++ clustering. Importantly, this does not imply that minimising quantisation MSE directly minimises the global training objective $F(\mathbf{w})$. Rather, it shows that MSE-optimal quantisation minimises a conservative upper bound on the additional loss induced by quantisation under worst-case smoothness assumptions.

\section{Curvature-Informed Quantisation Codebook}
\label{app:hessian_quant}

Appendix~\ref{app:kmeans_quant} used an $L$-smooth upper bound, which replaces the local curvature of the loss with a single worst-case scalar $L$. In this appendix, we relax the worst-case assumption and instead use the Hessian as a higher-order statistic to capture the current local curvature of the loss. The aim here is to motivate why a curvature-weighted approach might improve robustness. Note that, due to huge computational and rate overhead, computing the true Hessian (or even a reliable empirical Fisher in an FL setting) is infeasible, so the implementation tested in this paper uses a diagonal statistical proxy.

Reusing the notation and variables in Appendix~\ref{app:kmeans_quant}, we define the ideal and quantised model updates as
\begin{align}
    \mathbf{w}^{+} &=\mathbf{w}-\eta\bar{\mathbf{s}},\\
    \tilde{\mathbf{w}}^{+} &=\mathbf{w}-\eta(\bar{\mathbf{s}}+\mathbf{e}).
\end{align}

\subsection{Local Second-Order (Hessian) Demonstration}
Instead of applying the $L$-smooth descent lemma, we consider a local second-order expansion of $F$ around $\mathbf{w}$:
\begin{equation}
F(\mathbf{w}-\eta\mathbf{v})
\;\approx\;
F(\mathbf{w})
-\eta\nabla F(\mathbf{w})^\top \mathbf{v}
+\frac{\eta^2}{2}\mathbf{v}^\top H(\mathbf{w})\,\mathbf{v},
\end{equation}
where $H(\mathbf{w})=\nabla^2F(\mathbf{w})$ is the Hessian at $\mathbf{w}$. Applying this with $\mathbf{v}=\bar{\mathbf{s}}$ and $\mathbf{v}=\bar{\mathbf{s}}+\mathbf{e}$ and taking the difference, gives
\begin{equation}
\begin{aligned}
\Delta F
&\triangleq F(\mathbf{w}^+) - F(\mathbf{w}^-) \\
&\approx -\eta \nabla F(\mathbf{w})^\top \mathbf{e}
+ \frac{\eta^2}{2}\Big(
(\bar{\mathbf{s}}+\mathbf{e})^\top H(\mathbf{w})(\bar{\mathbf{s}}+\mathbf{e}) \\
&\hspace{3.2em}
- \bar{\mathbf{s}}^\top H(\mathbf{w})\bar{\mathbf{s}}
\Big). \\
&= -\eta \nabla F(\mathbf{w})^\top \mathbf{e}
+ \eta^2\,\bar{\mathbf{s}}^\top H(\mathbf{w})\mathbf{e}
+ \frac{\eta^2}{2}\,\mathbf{e}^\top H(\mathbf{w})\mathbf{e}.
\end{aligned}
\end{equation}
Using the same assumptions as in Appendix~\ref{app:kmeans_quant}, we have
\begin{equation}
\mathbb{E}[\mathbf{e}]=\mathbf{0},
\quad
\mathbb{E}\!\big[\nabla F(\mathbf{w})^\top\mathbf{e}\big]\approx 0,
\quad
\mathbb{E}\!\big[\bar{\mathbf{s}}^\top H(\mathbf{w})\mathbf{e}\big]\approx 0,
\end{equation}
which gives us
\begin{equation}
\mathbb{E}[\Delta F]
\;\approx\;
\frac{\eta^2}{2}\,\mathbb{E}\!\big[\mathbf{e}^\top H(\mathbf{w})\mathbf{e}\big].
\end{equation}
Compared to Appendix~\ref{app:kmeans_quant}, the key change is that the effective quantisation penalty and therefore the objective function to be used for constructing the quantisation codebook is now curvature-weighted. This implies that in areas where $H(\mathbf{w})$ has a larger magnitude, the same Euclidean error $\mathbf{e}$ contributes more degradation than in flatter loss-curvature directions. 

For the curvature-aware case, an extra step is needed since $H(\mathbf{w})$ is not necessarily positive semidefinite when away from a local minimum, and so $\mathbf{e}^\top H(\mathbf{w})\mathbf{e}$ is not guaranteed to be non-negative. To tackle this and reach a more useful objective, we replace $H(\mathbf{w})$ with a non-negative diagonal proxy
\begin{equation}
D(\mathbf{w})=\mathrm{diag}(d_1,\ldots,d_W),\qquad d_i\ge 0,
\end{equation}
allowing us to capture each coordinate's sensitivity in a positive semidefinite form. This leads to the surrogate objective
\begin{equation}
\begin{aligned}
   \mathbb{E}[\Delta F]
&\;\approx\;
\frac{\eta^2}{2}\,\mathbb{E}\!\big[\mathbf{e}^\top D(\mathbf{w})\mathbf{e}\big]\\
&\;=\;
\frac{\eta^2}{2}\,\mathbb{E}\!\big[\|D(\mathbf{w})^{1/2}\mathbf{e}\|_2^2\big]. 
\end{aligned}
\end{equation}

The result of this is that minimising the curvature-aware penalty corresponds to first re-scaling the BS local training output (which clustering is applied to) by $D(\mathbf{w})^{1/2}$, performing clustering in this transformed space, and then mapping the centroids back to the original space.

\section{Majority Vote and Trimmed Mean Methodology}
\label{app:nonlinear_aggregation}
Below, we outline the exact majority voting and trimmed mean aggregation rules used in Fig.~\ref{fig:aggregate_comp}, and clarify how they operate on the sparse count vectors recovered at the decoder.

For reference, the arithmetic-mean reconstruction (FedAvg-style baseline)  uses all recovered counts:
\begin{equation}
\hat{\mathbf{u}}^{\text{avg}}_i
=
\frac{1}{\hat{K}}
\sum_{j=1}^{n}\hat{x}_{i,j}\,\mathbf{q}_j,
\end{equation}
where $\hat{\mathbf{x}}_i\in\mathbb{N}^n_{0}$ is the recovered count vector in fragment slot $i$, $\hat{K}=\sum_{j=1}^n \hat{x}_{i,j}$, and $\mathbf{q}_j\in\mathbb{R}^d$ is centroid $j$ from the quantisation codebook.

\subsection{Majority Voting} 
Majority voting selects the most frequently used codeword in each fragment slot: 
\begin{equation}
\mathcal{T}_i \triangleq \arg\max_{j}\ \hat{x}_{i,j}. 
\end{equation}
If there is a unique winner ($|\mathcal{T}_i|=1$), we set $\hat{\mathbf{u}}^{\text{maj}}_i=\mathbf{q}_{j^\star}$ for the winning index $j^\star$. If there is a tie ($|\mathcal{T}_i|>1$), we break ties by averaging the tied centroids: 
\begin{equation}
\hat{\mathbf{u}}^{\text{maj}}_i = \frac{1}{|\mathcal{T}_i|} \sum_{j\in\mathcal{T}_i}\mathbf{q}_j. 
\end{equation}
This produces deterministic behaviour while discarding minority codewords, which is beneficial under sparse corruption, where corrupted devices tend to populate low-count indices, since normal device outputs are likely to collide, whereas quantised corrupt messages effectively act as random selections with a much lower chance of agreeing with other devices.

\subsection{Trimmed Mean}
Our trimmed-mean rule retains only the most common codewords until a target fraction of the device mass is reached. Let $\tau=0.8$ and define the retained mass as
$M \triangleq \lceil \tau \hat{K} \rceil$.
Indices are sorted by descending counts $\hat{x}_{i,j}$ and accumulate counts until the cumulative sum reaches $M$. In the simple case where $M$ is reached exactly, each index's vector is multiplied by its estimated magnitude, and the results are summed before being divided by $M$. The second case occurs when the cumulative sum exceeds $M$ at the final included index. In this case, only a fractional contribution of that index is used so that the total retained mass equals exactly $M$. Concretely, if the cumulative mass before index $j$ is $M_{\text{prev}}<M$ and adding $\hat{x}_{i,j}$ would overshoot, we assign that index a weight 
$w_j = M - M_{\text{prev}}$,
and stop the accumulation. The same magnitude-scaled averaging is then performed here. The third case is when there is a tie at the cut-off. If multiple indices share the same count at a given rank in the sorted list, we treat them as a group (no change). If the cumulative sum reaches $M$ exactly and the final included rank is a tied group, we include all tied indices with their full counts (since they fit exactly, there is no change). 
The only subtle case arises when the cutoff occurs \emph{inside} a tied group. In that situation, we split the remaining mass evenly across the tied indices. For example, suppose the next three indices are tied with counts $3,3,3$, but we have only mass $5$ left before reaching $M$. We average the 3 index's corresponding vectors, scale by the remaining mass $5$, and this contribution joins the cumulative sum to be scaled by $M$. This tie-handling keeps the trimmed mean deterministic and fair, while still discarding low-count (often corrupted) codewords. 



\bibliographystyle{IEEEtran}
\bibliography{main}

\end{document}